\documentclass[11pt]{article}

\makeatletter
\let\@fnsymbol\@arabic
\makeatother

\usepackage{cite} 
\usepackage{amsmath,amssymb,amsfonts}
\usepackage{algorithmic}
\usepackage{graphicx}
\usepackage{textcomp}
\usepackage{caption}
\usepackage{subcaption}
\usepackage{enumitem}
\usepackage[colorlinks=true,allcolors=blue,breaklinks=true]{hyperref}
\usepackage{breakurl}

\usepackage[centering]{geometry}    
\newcommand{\tsd}{360$^\circ$ }

\def\beq        {\begin{equation}}
\def\eeq        {\end{equation}}

\makeatletter
\newlength \figwidth
\if@twocolumn
\else
\fi
\makeatother

\begin{document}

\title{\vspace{-1.5cm}Towards Enabling Next Generation Societal Virtual Reality Applications for Virtual Human Teleportation\vspace{-0.3cm}\footnote{This is an extended version (with more details) of a tutorial feature article that will appear in the IEEE Signal Processing Magazine in September 2022.}}

\author{Jacob Chakareski\textsuperscript{$\star$}, Mahmudur Khan\textsuperscript{$\star,\dagger$}, and Murat Yuksel\textsuperscript{$\diamond$}\vspace{0.1cm}\\
{\small \textsuperscript{$\star$}Department of Informatics, New Jersey Institute of Technology,}\\ {\small \textsuperscript{$\dagger$}Department of Electrical and Computer Engineering, York College,}\\
{\small \textsuperscript{$\diamond$}Department of Electrical and Computer Engineering, University of Central Florida}}

\date{}

\maketitle

\begin{center}
\end{center}

\begin{abstract}
Virtual reality (VR) is an emerging technology of great societal potential. Some of its most exciting and promising use cases include remote scene content and untethered lifelike navigation. This article first highlights the relevance of such future societal applications and the challenges ahead towards enabling them.
It then provides a broad and contextual high-level perspective of several emerging technologies and unconventional techniques and argues that only by their synergistic integration can the fundamental performance bottlenecks of hyper-intensive computation, ultra-high data rate, and ultra-low latency be overcome to enable untethered and lifelike VR-based remote scene immersion. A novel future system concept is introduced that embodies this holistic integration, unified with a rigorous analysis, to capture the fundamental synergies and interplay between communications, computation, and signal scalability that arise in this context, and advance its performance at the same time. Several representative results highlighting these trade-offs and the benefits of the envisioned system are presented at the end.
\end{abstract}
\setcounter{footnote}{1} 

\section{Introduction and Motivation}
\label{sec:intro}
Virtual reality (VR) holds tremendous potential to advance our society. It enables visual immersion in virtual worlds created by means of computer graphics on a head-mounted display worn by a user and has found applications so far in training, education, entertainment, and gaming. An even broader set of use cases is anticipated ahead.



\begin{figure}[htb]
\centering
   \includegraphics[width=1.5\figwidth]{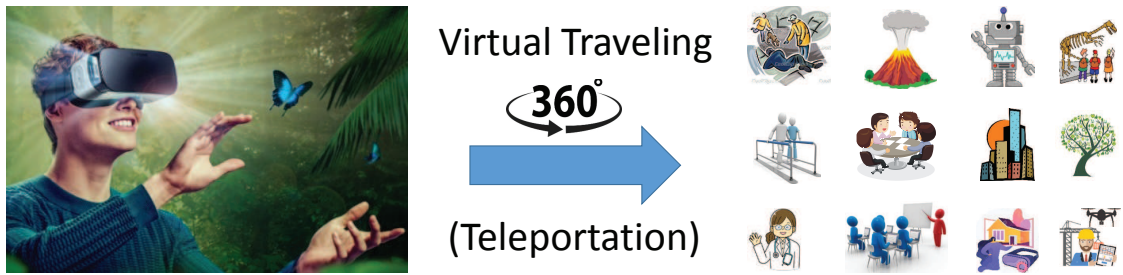}
   \caption{Seamless untethered virtual reality and lifelike remote scene \tsd video for virtual traveling/teleportation, to advance first responders, environmental monitoring, remote robot navigation and teleoperation, education and training, collaborative work, healthcare and rehabilitation, urban planning, and large-scale infrastructure inspection.}
   \vspace{-0.2cm}
   \label{fig:VR_360_Applications}
\end{figure}

Looking forward, VR is expected to make impact on quality of life, energy conservation, and the economy \cite{DigiCapital:16a,ApostolopoulosCCKTW:12}, and reach a \$62B market by 2027 \cite{GrandViewResearch:20}. As the Internet-of-Things (IoT) is becoming a reality, modern technologists envision transferring remote contextual and environmental immersion experiences as part of an online VR session. In particular, together with another emerging technology, known as 360$^\circ$ video, VR can suspend our disbelief of being at a remote location, akin to {\em virtual human teleportation} -- a truly momentous advance for our society \cite{Chakareski:18a}. The present state of the world (online classes, work from home, telemedicine, and so on) due to the COVID-19 pandemic aptly illustrates the importance of remote 360$^\circ$ video VR immersion and communication, enabled in a seamless, untethered, and lifelike manner across the spectrum of our society, as illustrated in Figure~\ref{fig:VR_360_Applications}.

However, two main highly-intertwined communication system challenges stand in the way of realizing this vision: VR requires {\bf (1) hyper-intensive computation and (2) ultra-low latency gigabit-per-second wireless networking}. Neither of these challenges can be met by current and upcoming conventional network systems \cite{Forbes_VR_2016,EdwardKnightlyKeynoteINFOCOM2017}, as the content to be delivered is too voluminous and the VR headsets' computing and storage capabilities are insufficient within an acceptable and wearable form factor. For instance, MPEG recommends a minimum of 12K high-pixel-quality spatial resolution and 100 frames per second temporal display rate, for a 360$^\circ$ video panorama experienced by a VR user \cite{ChampelSFTK:16}. These requirements would map to a data rate of several Gbps, even after applying state-of-the-art High Efficiency Video Coding (HEVC) compression \cite{SullivanOHW:12}. Similarly, mobile GPUs lag their desktop counterparts in computing power by a factor of ten and will not have the required TFlops to provide the necessary VR decoding and rendering computation, at the minimum resolutions and frame rates indicated above, in the foreseeable future, given the current semiconductor technology trends \cite{CuervoCK:18}.

Emerging 360$^\circ$ video practices compound these challenges, as they are highly inefficient and reuse traditional video/networking technologies in virtual reality contexts without integrating their specifics \cite{Petrangeli2017}. This considerably degrades the quality of experience and application utility. The under performing of these practices is even more dramatic in mobile settings, due to the much lower wireless bandwidth and computing capability of such devices. Regrettably, this is the context where such applications are expected to have the highest societal impact, {\em advancing disaster relief, the environmental sciences, public safety, transportation, search and rescue, and urban planning}, among others.

Therefore, these considerable challenges and shortcomings limit present VR applications to off-line operation, low-fidelity graphics content, tethered high-end computing equipment, and predominantly gaming and entertainment settings.

\section{Objectives of the Feature Article}
This article has multiple objectives of educating the broader Signal Processing Magazine readership. Its first aim is to educate about the importance of enabling next generation virtual reality applications comprising high-fidelity remote scene immersion and seamless untethered lifelike navigation of the reconstructed remote environment. Its second aim is to emphasize that traditional technology upgrade cycles alone would not suffice to bridge the performance gap to make such applications possible and that a holistic integration of unconventional techniques and emerging technologies would be required instead. Its third aim is to provide a tutorial of these methods and their synergistic interplay towards enabling the envisioned next generation applications. Finally, its fourth aim is to illustrate a case study of rigorous high-level integration of these strategies and systematic end-to-end modeling and analysis into an embodiment of a future mobile multi-user virtual reality system for six degrees of freedom (6DOF) immersion. Besides its educational objectives, a fifth aim of the feature article is to identify a research framework and stimulate novel research and community building, given the emerging nature of virtual reality and 360$^\circ$ video and their prospective broad societal impact.

Another benefit of the fifth aim is that the technical advances it can lead to can facilitate fundamental research in other application areas of high-volume high-speed/low-latency data transfer in emerging cyber-physical systems and IoT settings, where for the first time the spatiotemporal aspects of the data need to be closely explored and tightly integrated with the user navigation actions, to maintain the desired quality of experience for the end user, given the limited available system resources.

\section{Existing Tutorials and Distinctions}
There have been tutorials appearing before at conferences such as IEEE VR and 3D User Interfaces, and ACM SIGGRAPH. However, they focused on unrelated aspects such as eye tracking in 360-degree video, human perception in virtual environments, and virtual reality content creation \cite{LeMeurJ:19,GabbardSE:12,Isdale:03}, which are traditionally associated with the fields of computer vision and human perception. Similarly, another recent brief article from the IEEE Communications Magazine \cite{BastugBMD:17} narrowly focuses on 5G (a technology upgrade cycle) and an abstract wireless VR application as one prospective use case, using assumptions that do not relate well to practice and thus do not lead to insightful observations. These include, for instance, traditional models of user arrival processes from communications/information theory, immersive experience measures expressed in percentages, and unrealistic compression rate characteristics of 4K \tsd video content.
The scope of this feature article is very different and broader, as it aims to highlight and educate about the present challenges of enabling future virtual reality applications, deployed in an untethered manner and with high-fidelity remote scene content, focusing on fundamental problems and trade-offs between signal processing, communications, and computation that arise in this context. Another objective of the article is to provide a tutorial like coverage of a non-conventional framework of research that can help bridge the present performance gap to enable such future applications and harness their expected societal benefit.



\section{Progress to date, synergistic advances, and broader implications}

We outline related work in \tsd video VR streaming using traditional approaches and systems, and synergistic advances in other technological domains that can help overcome the fundamental bottlenecks of the former.
Relative to traditional video streaming  \cite{ChakareskiF:06,ChakareskiAWTG:04a,ReisCKS:10,ChouM:02}, 360$^\circ$ video streaming to VR headsets introduces the additional challenging requirements of ultra-high data rate, hyper-intensive computation, and ultra-low latency, as introduced earlier. Though some advances have been made in \tsd video streaming using traditional network systems, by investigating intelligent resource allocation and content representation \cite{Petrangeli2017,ChakareskiACSS:17,CorbillonDSC:17,HosseiniS:16}, the delivered immersion is still limited to low to moderate quality and 4K spatial panorama resolution, encoded at a temporal rate of 30 frames per second. This outcome stems from the fundamental limits in data rate and latency of such systems and their use of traditional server-client architectures. Moreover, a shared key shortcoming of the majority of emerging studies that is important to note is the pursuit of heuristic design choices and the lack of analysis of the fundamental performance trade-offs among the delivered immersion quality, user navigation patterns, signal representation, and system resources.



Free-space optics (FSO) and millimeter wave (mmWave) are emerging wireless technologies that are presently investigated and developed to help overcome the bottlenecks of traditional wireless systems. Both have the potential to enable multi-Gbps data transmission rates. FSO exploits the light intensity of a light emitting diode (LED) or a laser diode (LD) to modulate a message signal. FSO technologies using the former approach are known as visible light communication or VLC, as they provide illumination at the same time. After propagating through the optical wireless channel, the (infrared or visible) light message is detected by a photo-diode (PD) \cite{dimitrov2015principles}.
Unlike the radio frequency spectrum, plentiful unlicensed spectrum is available for light communications (300GHz–800THz), which has put FSO on the road-map towards sixth generation (6G) networks \cite{8792135}. While being a novel technology, a few studies of design concepts and experimental testbeds have already appeared \cite{rahman2018fso,8998135}. In the radio frequency spectrum, mmWave wireless communication is considered the enabling technology of next-generation wireless systems, as in the range of 10-100 GHz, more than 20 GHz of spectrum is available for use by cellular or wireless LAN applications. mmWave has seen its first commercial products operating in the 60~GHz band appear in the early 2010s. More complex transmission schemes to increase even further the achievable data rate are currently being investigated \cite{blandino2018multi}.

A few disparate preliminary studies emerged so far examining the potential of FSO and mmWave to advance mobile VR. A mmWave-based VR system was proposed in \cite{LiuZZLZZG:18} that uses WiGig modules for wireless connectivity between two laptop computers, one acting as a server and the other as a client, with the VR headset attached to it. Similarly, a mmWave reflector was developed in \cite{AbariBDK:17} to aid in connectivity maintenance with a mobile VR headset in the event of blocking of the direct (line of sight) wireless link. In each case, only synthetic computer graphics content was considered for transmission to and rendering on a VR headset. The company HTC released recently a wireless adapter operating at 60 GHz to enable untethered delivery of computer graphics content to VR headsets. However, the quality of experience is limited, due to the low quality and low resolution of the delivered content, and the highly ineffective real-time compression applied at the transmitting server. A design concept for using narrow beam FSO transceivers was proposed in \cite{rahman2018fso}, highlighting the challenge of link maintenance in such settings. The study in \cite{khan2019visible} explored the design of an FSO-enabled VR headset featuring hemispherically organized layers of highly directional PDs to facilitate connectivity maintenance under challenging head navigation movements of a VR user. Finally, Microsoft patented last year a design concept for a free space optics enabled VR headset and transmitting system \cite{CuervoGCK:19}, indicating vision and interest in the tech sector for actual products in the future.

In earlier related developments highlighting the prospective benefits of raw video data transmission, delivery of traditional uncompressed HD video over a short range 60 GHz wireless link was studied for home settings targeting stationary consumer electronics that do not integrate video compression, e.g., gaming consoles \cite{SinghQSNKK:08}. More recently, due to its ability to reduce network latency and help mobile devices offload part of their computation, edge computing and caching have started to be investigated for delivering \tsd content in wireless cellular systems \cite{Chakareski:18}.

The studies and endeavors highlighted heretofore either address specific technology aspects or lack a systematic end-to-end analysis. The envisioned next generation VR system 
can help broadly advance the above efforts and the state-of-the-art, and make simultaneous impact on other emerging application areas of similar characteristics, as noted earlier. 
Our preliminary advances highlight the substantial benefits and potential of the envisioned system \cite{ChakareskiK:21,ChakareskiG:20}.

As complementary to the above discussion, we examine if technologies employed in online multi-player VR gaming could be leveraged to enable further benefits. Foveated rendering exploits the very narrow field of view (the central $1.5-2^\circ$ of the entire field of view of the human eye) of the fovea, responsible for sharp central vision, to reduce the rendering workload of generating an image to be displayed on a VR headset, by greatly reducing its quality in the peripheral vision (outsize the zone gazed by the fovea) \cite{FoveatedRenderingWikipedia:21}. Essentially, much fewer pixels and at lower fidelity are rendered outside the fovea area. This considerably accelerates the frame rate of display of successive image frames displayed on a headset, which is helpful in dynamic gaming environments. Foveated rendering can be integrated into the envisioned VR system towards the same goal. Still, two aspects need to be considered carefully. Gaming content can be degraded considerably without noticing artifacts up close, as the computer graphics content is not overly complex and is evolving rapidly. On the other hand, high fidelity remote immersion via \tsd video may be susceptible to visibly noticeable degradation in quality under the same setting, as actual scene content is much more complex (to render) and thus sensitive to pixel resolution or fidelity reductions across the entire field of view. Even more importantly, the most challenging computing task faced by a headset in \tsd video VR applications is decoding the massive content at the target frame rate, not rendering the user's viewpoint, as explained later.

Online gaming applications also facilitate server-client architectures featuring multiple distributed servers to handle the massive client load \cite{BharambePS:06}. Each server maintains and updates a copy of the shared gaming environment, in response to the players' actions controlling their respective avatars in the game. An updated state of the environment is then sent back to all players every 100ms for first person shooter games (most dynamic) or at a lower rate for other types of games. Each player updates its own view of the environment using its gaming device based on the received updated state. Though such traditional computer network systems are suitable for this type of applications, the latencies, data rates, and weak state inconsistencies they exhibit would not be conducive to enable lifelike remote scene immersion on untethered VR headsets.



\section{A Virtual Reality and \tsd Video Primer}

\begin{figure}[htb]
\centering
   \includegraphics[width=\columnwidth]{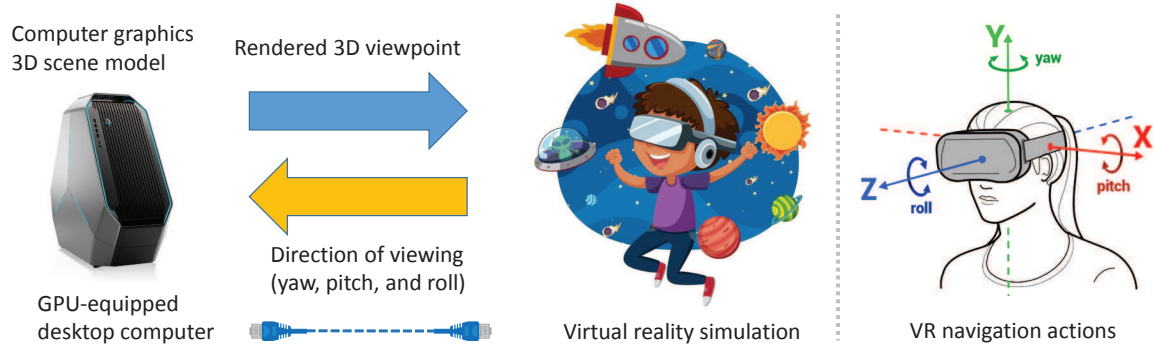}
   \caption{VR 101: {\bf (Left/Middle)} A user is linked to a powerful (gaming) desktop computer, equipped with a GPU and computer graphics software, to experience a VR simulation on her headset. The rendered 3D viewpoint of the user in the simulation and the direction of viewing of the user are exchanged over a long high data rate multimedia cable. {\bf (Right)} The navigation actions of the user comprise rotation angles yaw, pitch, and roll around three coordinate axes centered on the user's head and determine the direction of viewing.}
   \vspace{-0.2cm}
   \label{fig:VR_Primer}
\end{figure}

A high level system illustration of a VR application is included in Figure~\ref{fig:VR_Primer}, left and middle. A user wearing a VR headset is linked to a powerful (typically gaming) computer, equipped with a Graphics Processing Unit (GPU), computer graphics software, and a 3D scene model. The link comprises a thick long integrated cable featuring as its main component a high data rate multimedia cable such as HDMI that transports in the direction of the user her present 3D viewpoint in the VR simulation rendered by the server computer based on the direction of viewing of the user communicated over the cable in the opposite direction. The navigation actions of the user comprise three head rotation angles denoted as yaw, pitch, and roll, and are illustrated in Figure~\ref{fig:VR_Primer} right. They are measured with respect to three coordinate axes centered on the user's head, using gyroscopes built into the VR headset, and determine the direction of viewing of the user in the synthesized 3D scene.

The virtual scene may be static or dynamically evolving. The simulation may also include spatial audio content that is reproduced in parallel on stereo headphones worn by the user. More recent VR application systems include the possibility for limited spatial movement of the user in the virtual scene. The spatial coordinates of the user headset are then tracked externally using infrared base stations mounted in the room where the system resides. The present 3D viewpoint of the user is rendered based on her spatial coordinates and direction of viewing in this case. 

The major computing load of the application is reconstructing the 3D viewpoint of the user dynamically in response to her navigation actions and can be quite intensive. It is handled by the server computer and its GPU that execute demanding computer vision algorithms on the voluminous geometric representation of the 3D virtual scene. The collocated server computer and VR headset, and the cabled high data rate connection between them help to minimize the interactive latency of the application and to deliver high volume 3D computer graphics at the display frame rate required to avoid motion sickness \cite{Moss2011}.

Still, having a tethered VR headset can represent a tripping hazard and reduces the quality of experience of the user and the utility/scope of the application. Thus, most recent VR application systems feature wireless headsets, with stand alone display and computing capabilities or a slot in which a mobile phone is inserted to provide them. However, the capabilities of such systems in terms of delivered content quality, frame rate, and interactive latency are not on par with their tethered counterparts, as described earlier.





\begin{figure}[htb]
\centering
   \includegraphics[width=\figwidth]{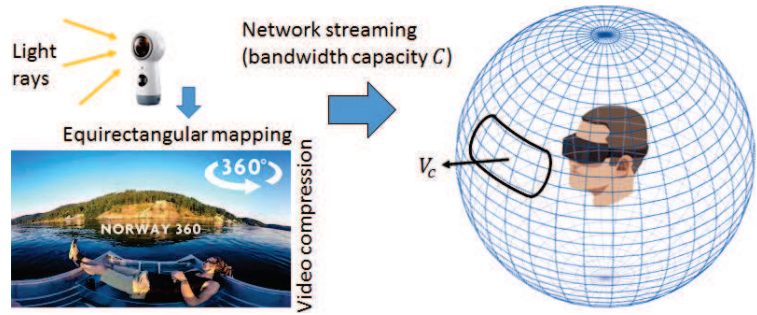}
   \caption{360$^\circ$ video capture and streaming, and user viewport $V_c$.}
   \vspace{-0.2cm}
   \label{fig:360VideoBackground}
\end{figure}

A \tsd video VR application system replaces the computer graphics 3D scene model from Figure~\ref{fig:VR_Primer}, left and middle, with actual remote scene 3D content. In particular, 360$^\circ$ video is a recent video format that is recorded by an omnidirectional camera that captures incident light waves from every direction (see Figure~\ref{fig:360VideoBackground}, top left). Such cameras comprise two or more wide-field lenses and rely on computer vision image stitching to produce a full 360$^\circ$ horizontal plane field of view and 180$^\circ$ vertical plane field of view \cite{Omni360CameraWikipedia:20}. Thereby, the constructed scene content would appear captured on the interior surface of a sphere centered on the camera location, akin to how we perceive the world around us.

%

Concretely, \tsd video enables a 3D 360-degree look-around of the surrounding scene for a remote user, virtually placed at the camera location, on his VR headset, as illustrated in Figure~\ref{fig:360VideoBackground} right. After capture, the raw spherical or 360$^\circ$ video frames are first mapped to a wide equirectangular panorama (illustrated in Figure~\ref{fig:360VideoBackground}, bottom left) and then compressed using state-of-the-art (planar) video compression such as HEVC. The former intermediate step is introduced, as compression techniques operating directly on spherical data are much less mature and performing relative to traditional video compression operating on 2D video frames. Beyond the equirectangular mapping, which is most widely adopted, cube, pyramid, and dodecahedron planar projections have also been studied \cite{CorbillonDSC:17}.


The computing workload of a \tsd video VR application system is even more intensive and comprises decoding the compressed \tsd content and rendering the current 3D viewpoint of the user dynamically in response to his navigation actions. The latter is informally known as viewport and corresponds to only a small portion of the 360$^\circ$ view sphere denoted as $V_c$ in Figure~\ref{fig:360VideoBackground} right. The task of reconstructing the viewport on the VR headset requires remapping the decoded 360$^\circ$ panorama to the original spherical format and then projecting pixels from $V_c$ to their planar equivalents on the display of the headset.

For remote service, when the user and the stored 360$^\circ$ data are not collocated, the entire monolithic 360$^\circ$ panorama is commonly streamed to the user using traditional state-of-the-art video streaming (MPEG-Dynamic Adaptive Streaming over HTTP (DASH) \cite{Sodagar:11}). This considerably penalizes the quality of experience, due to the overwhelming volume of 360$^\circ$ data that needs to be delivered and that exceeds by orders of magnitude the available network bandwidth $C$ and the computing capabilities of the receiving device. These shortcomings are compounded by the reliance on ineffective network protocols such as HTTP and TCP to deliver latency-sensitive multimedia data of this nature, which have been adopted to lower the cost of an intelligent streaming system at the penalty of not having good control of the data delivery process. Thus, only lower quality, frame rate, and resolution 360$^\circ$ videos can presently be delivered online over the Internet. Yet, the streaming also lacks the ultra-low latency interactivity needed for truly immersive experiences, as traditional server-client Internet architectures are used in this case. The quality of experience and application utility are even lower in traditional wireless settings, due to the even lower data rates and computing/storage capabilities available therein, as noted earlier.

It should be noted that the recent advances in spatially adaptive streaming of broad or omnidirectional video panoramas \cite{CorbillonDSC:17,Petrangeli2017,ChakareskiACSS:17,HosseiniS:16} have contributed to further activities within MPEG and the follow-up DASH-SRD and OMAF standards that integrate them \cite{MPEG-DASH-SRD:15,MPEG-DASH-OMAF:18}. However, most of the bottlenecks highlighted above of such standards-based traditional approaches and their fundamental limitations in terms of enabled data rate, delay, and computing capabilities still remain.

\section{Bridging the Present Performance Gap}
\begin{figure}[htb]
\centering
   \includegraphics[width=1.2\figwidth]{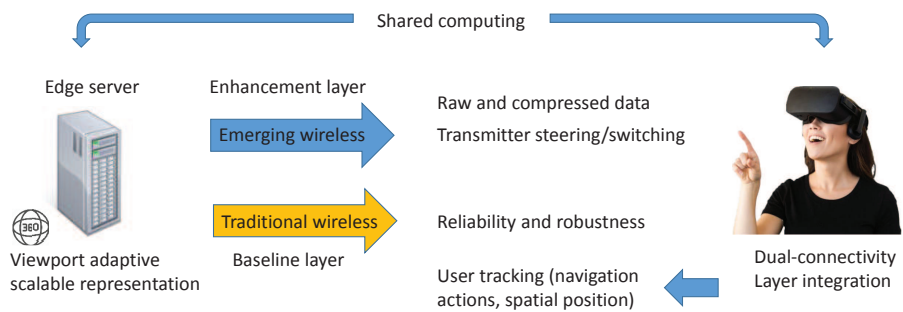}
   \caption{Future 6DOF virtual reality system to bridge the present performance gap. It integrates synergistic edge computing, viewport-adaptive scalable \tsd video representation, and dual-connectivity transmission via traditional/emerging wireless technologies.}
   \vspace{-0.2cm}
   \label{fig:NextGenerationVRSystem}
\end{figure}

In the following, we provide a high-level tutorial-style description of several unconventional techniques and emerging technologies and their synergistic interplay towards enabling next generation virtual reality applications comprising high-fidelity remote scene immersion and seamless untethered lifelike navigation of the reconstructed remote environment. To set the discussion, we illustrate in Figure~\ref{fig:NextGenerationVRSystem} a novel 6DOF VR system concept that embodies their holistic integration to help bridge the present performance gap.

Its key components are an {\em edge server equipped with storage and computing capabilities} that will help bring the content and required computing closest to the user without overwhelming her device, to minimize the interactive latency; a {\em viewport-adaptive scalable \tsd video representation} that will facilitate dynamic spatiotemporal adaptation of the 6DOF content in response to the user's actions, to maximize the transmission efficiency; and {\em dual connectivity transmission via traditional and emerging wireless technologies} that will simultaneously provide reliability and robustness, and high-fidelity immersion. The dual-connectivity transmission will stream in parallel two synergistic content layers (baseline and enhancement) over the two respective wireless technologies. The enhancement layer will comprise raw and compressed content data to enable further performance synergies.

To enable its effective use, the transmission over the emerging wireless technology will integrate transmitter steering or switching that will facilitate user tracking feedback capturing the navigation actions of the user and her spatial position. The user will be equipped with a dual-connectivity VR device that will integrate the two content layers.

Given the nature of the application, it will be an effective system design to employ dual-connectivity transmission only for the down-link communication from the edge server to the mobile VR users. The up-link communication in the opposite direction, i.e., from the users to the edge server, can be effectively carried out by employing solely single-connectivity transmission via reliable lower data rate traditional wireless technology links, as it will only carry miscellaneous low-rate control information.

%
%
%


We implicitly have an indoor setting in mind for the envisioned system, due to the nature of the target application. Moreover, some of the emerging technologies and unconventional techniques the system integrates have advanced further and would be easier to deploy in such a setting. Still, all its key components can either directly apply to an outdoor scenario or have outdoor counterparts that have been developed in parallel. Thus, an outdoor deployment of the envisioned system can potentially be pursued as well.


\subsection{Raw data transmission of ultra-high resolution and frame rate content}
Video compression has made the present Internet possible and has fueled its growth for many years \cite{CiscoVNI_BarnettJAK:18}. It enables a reduction of the required network transmission data rate by a few orders of magnitude. However, it induces a decoding delay at the receiving client, at the same time, proportional to the compressed data volume. This can be penalizing for ultra-low latency applications such as online remote immersion via virtual reality, especially in mobile settings where client devices have limited computing capabilities and the delivered compressed data is massive. Intelligent transmission of the required 360$^\circ$ video content as raw data in this setting can help overcome this challenge, as explained next.

\subsection{High-frequency directional wireless transmission}
It may seem contradictory at first to send raw video data as that would scale back up the transmission delay, by a few orders of magnitude. However, the emerging wireless technologies of free-space optics and millimeter wave transmit data at much higher spectrum frequencies (relative to traditional sub-6 GHz wireless technologies), to enable several orders of magnitude higher data rates. Using such transmission can make sending raw video data appealing, as it would cancel out the increase in transmission delay and still provide the benefit of lower computing delay at the client device.


However, these emerging technologies exhibit different transmission characteristics relative to their traditional counterparts that need to be addressed to enable their efficient utilization. Concretely, their transmission beams are very narrow and directed, as reflected waves from obstacles and the environment feature very poor signal quality, due to the high career frequency that is used for transmission \cite{Popovski:20}. Therefore, the transmitter and receiver need to be positioned in a direct line-of-sight of each other and be actively aligned to maintain that property, in case of a mobile receiver. This challenge can be addressed in the envisioned system by benefiting from the available user tracking information and the limited spatial mobility of the users in an indoor setting.

In particular, high-frequency free-space optical transmitters can be mounted on steerable platforms and actively directed towards the users using servo motors, based on the tracking information. Two servo motors can control the horizontal and vertical rotations, or azimuth and elevation angles, of a transmitter tracking its user over the spatial area of the system. Such motors can operate at 50 Hz control signal frequency and can cover a \tsd rotation angle per second \cite{TowerProMG995ServoMotor:21}, which is much faster than a typical mobile user maximum speed of one meter per second indoor \cite{TavakkolniaSAGASH:19}. Similarly, high-frequency millimeter wave transmitters adequate for mobile communication can be realized with multi-element phased-antenna arrays (PAAs) to enable dynamic beam forming in both azimuth and elevation in response to the navigation/mobility actions of the respective receiving users \cite{blandino2018multi}.


%
VR headsets can be equipped with respective receivers for such high-frequency transmission.
In the case of free-space optics, the upper portion of a headset can be covered with an array of photo-diode detectors, which can receive and decode in parallel the optical beam signal incident on their surface. The decoded incident signals at every photo-detector can be combined using diversity combining techniques for more effective performance. The dimensions and arrangement of the photo-detector array can be configured to match the accuracy of the steering motors of the respective transmitters, the width of the optical beam, and the head movement navigation actions of a user. Similarly, a small form factor millimeter wave multi-element PAA receiver\footnote{A prototype of size $5 \times 20$ millimeters, featuring a 16-element PAA, was recently demonstrated in \cite{blandino2018multi}.} 
can be placed on the top portion of a VR headset. 
As the respective transmitters for each technology would be typically placed high above the users, blockage by human movement would not be a challenge and the placement of the receivers on top of the VR headsets would be a natural choice.

\begin{figure}[htb]
\centering
\begin{subfigure}{.5\columnwidth}
  \centering
  \includegraphics[width=0.9\linewidth]{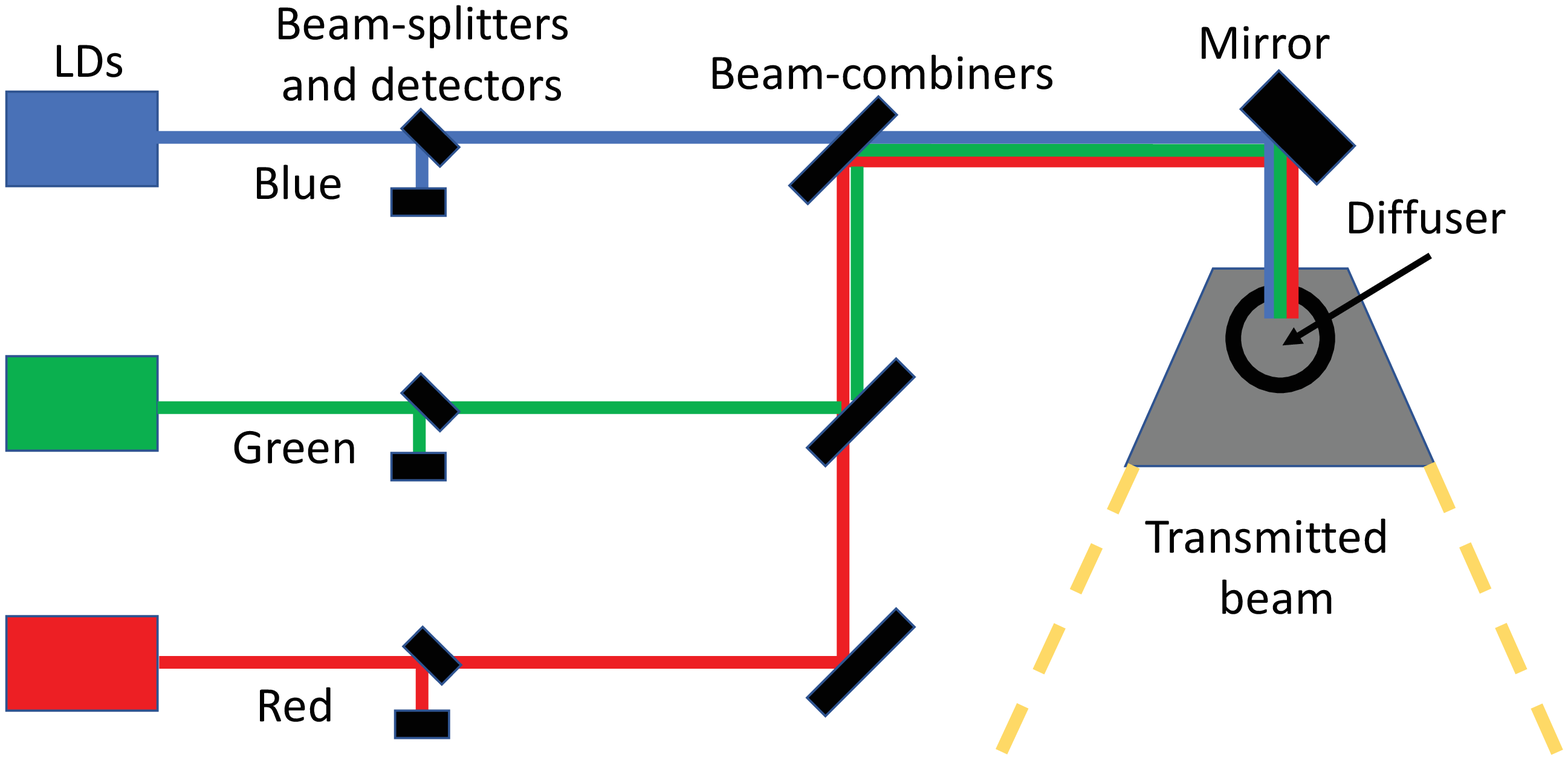}
\end{subfigure}%
\begin{subfigure}{.5\columnwidth}
  \centering
  \includegraphics[width=0.9\linewidth]{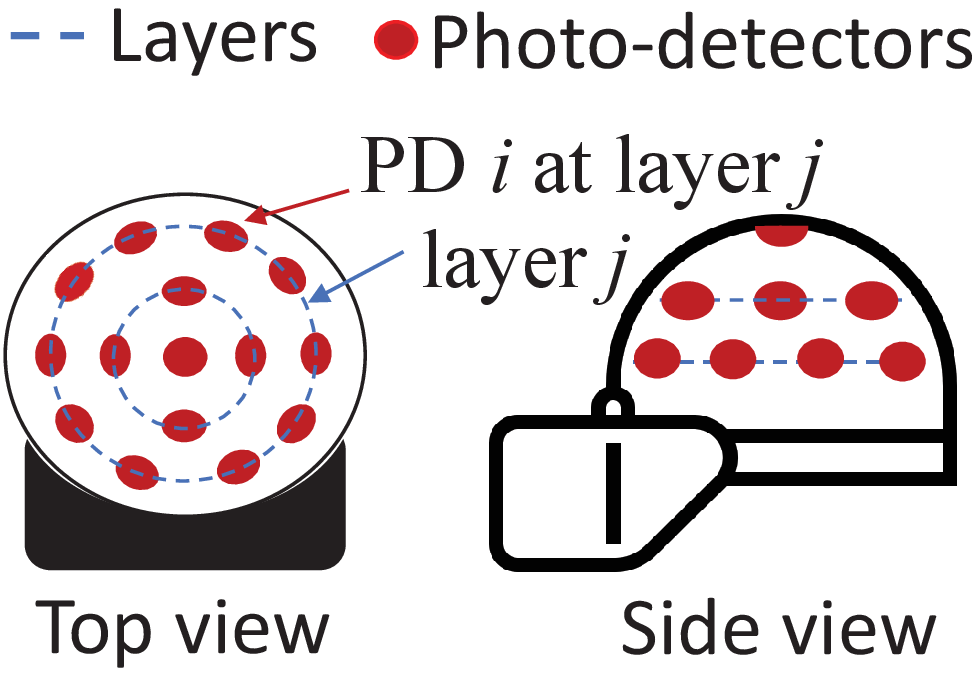}
\end{subfigure}
\vspace{-0.3cm}
\caption{RGB-LD transmitter (left) and VR headset equipped with photo-detectors (right).}
\label{fig:RGB_LD_VLC_Transmitter_VR_HeadsetPhotoDiodes}
\vspace{-0.3cm}
\end{figure}


We discuss in slightly more detail now an embodiment of the FSO transmitters and receivers in the envisioned system. LDs can enable higher transmission rates relative to LEDs, and thus may be the preferred choice. As their transmission beam is very narrow (typically $< 1^\circ$), a combination of red, green, and blue LDs, together with a diffuser, as illustrated in Figure~\ref{fig:RGB_LD_VLC_Transmitter_VR_HeadsetPhotoDiodes} (left), can be used to produce a slightly broader transmission beam that will increase the transmission coverage and reliability, while limiting the reduction in data rate. Simultaneously, this will lead to coherent white light emitted by the transmitter that can prospectively serve as lighting. Early prototypes of such transmitters have demonstrated data rates of up to 10 Gbps \cite{zafar2017laser}. Beam splitters and detectors can be used to monitor and control the different laser powers, for safety/performance considerations.



To mitigate the effect of narrow transmission beams and maintain high receive data rate, an angle-diversity-receiver can be implemented, featuring multiple small-area PDs installed on a platform at different inclination angles\footnote{The receive data rate of a PD rapidly drops with its surface area. However, a large surface area receiver increases the reliability of the optical link. The envisioned receiver design integrates the best of both approaches.}. Such a multi-PD design can be particularly helpful in maintaining the highly directional optical link during dynamic 360$^{\circ}$ navigation head movements of a VR user. We recently investigated this approach towards the design of a helmet-shaped VR headset, comprising multiple PDs placed along different upper-hemispherical layers on the headset, as illustrated in Figure~\ref{fig:RGB_LD_VLC_Transmitter_VR_HeadsetPhotoDiodes} (right). One PD is placed on top, and the rest are distributed along different layers on the hemispherical headset surface. The design optimization aims to minimize the number of PDs used, while ensuring uninterrupted connectivity with an FSO transmitter. It selects the number of layers, their placement along the headset hemisphere, and the number of PDs per layer, in pursuit of this objective, and integrates VR head movement navigation data for robust connectivity \cite{khan2019visible}. Power dissipation of the PDs is important in this setting and can be integrated into the design optimization, to impact the number of PDs placed on the headset.

\subsection{Low-high-frequency dual connectivity wireless streaming}
High-frequency wireless transmission induces brittle pencil-beam like directed communication links. Streaming 360$^\circ$ video over them alone would lower the reliability, as such links can be fragile and sensitive to line-of-sight misalignment between the transmitter and the receiver. Integrating low-frequency (traditional sub-6 GHz) wireless transmission as a synergistic supplement can help maintain the robustness of the system, as such radio waves do not need a sender-receiver alignment and provide multi-path and reflected signal benefits. Concretely, the traditional wireless connectivity can be employed to stream a baseline layer or representation of the 360$^\circ$ content that alone will ensure an uninterrupted service and reliable application quality, if a transient high-frequency link loss occurs prospectively. The high-frequency wireless connectivity can then be used to stream an enhancement content layer or representation that will build upon the baseline layer to enable high fidelity immersion and augment the quality of experience of the user. Due to the plentiful network transmission bandwidth enabled by the high-frequency connectivity, the enhancement content layer can be streamed at least in part as raw data, to enable further system performance advances. An effective operation of such dual-connectivity 360$^\circ$ video streaming towards reliable high-fidelity immersion can be advanced via a synergistic and efficient design of the data representation that will enable the construction of the two content layers, as explained next.

\subsection{Scalable 360$^\circ$ video tiling and viewport-driven adaptation}
\label{sec:Scal360TilingViewportAdaptation}
Rather than streaming the entire 360$^\circ$ panorama, as conventionally done, one can construct a scalable tiling-based representation of the 360$^\circ$ panorama to enable effective spatiotemporal adaptation of the transmitted data stream to the dynamic viewport of the user and the available network bandwidth. This will lead to efficient use of resources and higher quality of experience for the user. Moreover, the scalable 360$^\circ$ tiling can facilitate rigorous mathematical analysis of performance aspects of streaming systems, as we show later. Tiling of traditional wide-panorama video has been introduced as an option in HEVC, to facilitate parallel compression of the content in multi-core processor systems. More recently, its benefits have started to be recognized for 360$^\circ$ video streaming \cite{ChakareskiACSS:17,HosseiniS:16}.

\begin{figure}[htb]
  \centering
  \includegraphics[width=1\linewidth]{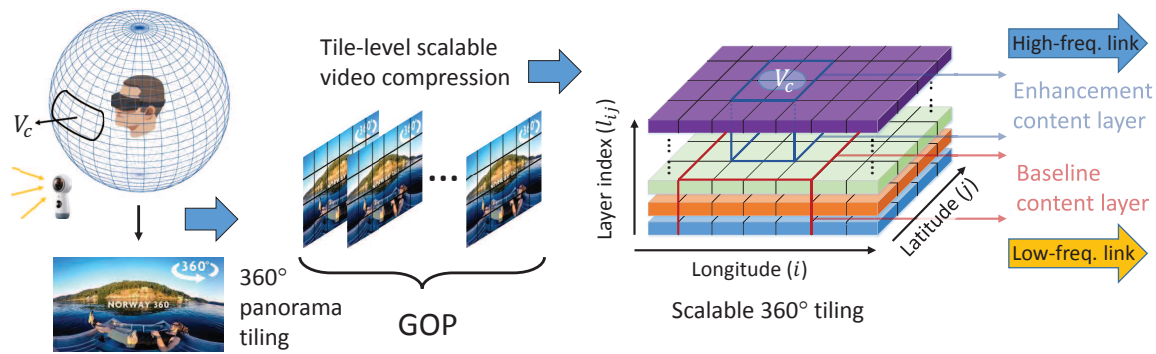}
  \caption{GOP-level scalable signal tiling of an equirectangular 360$^{\circ}$ panorama.}
  \vspace{-0.2cm}
  \label{fig:360VideoBackgroundTilingStreaming}
\end{figure}

A sample scalable tiling-based 360$^\circ$ video representation is illustrated in the right portion of Figure~\ref{fig:360VideoBackgroundTilingStreaming}. Concretely, each 360$^\circ$ video frame comprising a Group of Pictures or GOP\footnote{This is a block of consecutive frames that are compressed together with no reference to other frames.} is spatially partitioned into a set of tile sectors or simply tiles $(i,j)$ along its longitude and latitude dimensions. The collection of tiles across the video frames of a GOP at the same spatial location is then compressed into multiple embedded layers of progressively increasing signal fidelity. The first (bottom) layer of a compressed 360$^{\circ}$ GOP-tile is commonly known as a base layer, and the remaining layers are identified as enhancement layers. The reconstruction fidelity of a GOP-tile improves incrementally as more layers are being decoded progressively starting from the first layer.

The baseline content layer can be constructed from the scalable tiling-based 360$^\circ$ representation such that it comprises the first $L_b$ layers for a broader set of tiles encompassing the user viewport, to account for (i) a prospective mismatch between viewport knowledge at the server/sender, used to construct a complementary enhancement content layer, and the actual user viewport at the receiving client, induced by rapid user head movements, or (ii) a prospective transient high-frequency link loss. Accounting for these two possibilities thereby will ensure that the viewport can be reconstructed continuously and will augment the application reliability considerably. The enhancement content layer can then be constructed such that it comprises the subsequent $L_e$ layers for a narrower set of tiles focusing closely on the viewport, to maximize its expected quality, as illustrated in the right portion of Figure~\ref{fig:360VideoBackgroundTilingStreaming}. The choice of number of layers and tile selection from the scalable 360$^\circ$ tiling to construct the baseline and enhancement content layers can be made rigorous, inclusive of the integrated selection of a subset of enhancement content layer tile to be transmitted as raw data, as introduced earlier.
We illustrate this as part of an end-to-end example analysis of the envisioned VR system concept that we present later. 




%
%
%
%
%
%
%
%


We note that cheap storage and easier implementation have made non-scalable video compression preferred in practice. With minor adjustments, our system design can directly apply to the case of constructing both content layers independently using non-scalable compression. Moreover, scalable video coding has been successfully deployed in cutting-edge technologies for diverse low-latency multi-party telepresence settings \cite{Vidyo:22}, providing therein considerable benefits in terms of lower server complexity and higher client quality of experience over state-of-the-art non-scalable video compression based solutions. These benefits and related system context arise in the setting investigated here as well. Finally, scalable video coding is consistently enhanced through research and every subsequent generation of video codecs \cite{BoyceYCR:17}. The next generation VR system we explore in this feature article can provide avenues for advancing further such efforts.

\subsection{6DOF Virtual Reality and \tsd Video}
\begin{figure}[htb]
  \centering
  \includegraphics[width=0.85\linewidth]{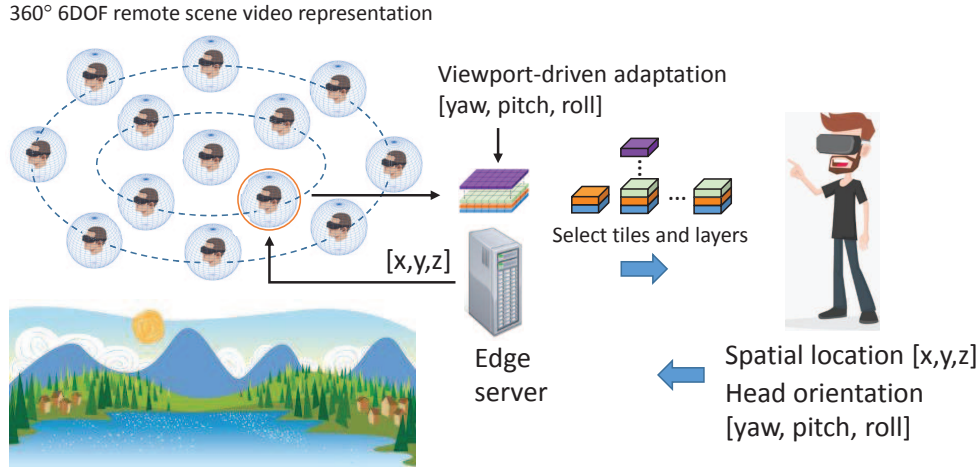}
  \caption{6DOF remote scene immersion, and dynamic spatial \tsd viewpoint and viewport adaptation for a mobile VR user navigating the scene.}
  \vspace{-0.2cm}
  \label{fig:6DOF_360VideoStreaming}
\end{figure}

A single \tsd video (of actual or synthetic content) enables three degrees of navigation freedom (3DOF) to a VR user, in the form of rotational head movements around three orthogonal axes (as noted earlier, see Figure~\ref{fig:VR_Primer} right), in experiencing a remote scene immersively from a single location. The streaming strategy described in Section~\ref{sec:Scal360TilingViewportAdaptation} can be extended naturally to the case when the application will also allow for spatial movement of the user in the remote scene, to enable a 6DOF immersion experience over the spatial area of the VR arena, where the content is navigated. 
Here, the user will have the ability to select not only her direction of viewing but also the spatial location of the \tsd viewpoint in the scene to be explored, which will augment the quality of experience. The content for each such spatial \tsd video viewpoint available to be navigated can be represented using the scalable \tsd tiling approach highlighted above and the edge server can apply in this case dynamic viewpoint adaptation and viewport-driven content adaptation jointly, in response to the 6DOF navigation actions of the user, as illustrated in Figure~\ref{fig:6DOF_360VideoStreaming}. 

Complementary signal processing explorations can be pursued in such a 6DOF system, for further enhancements. For instance, additional {\em virtual} \tsd video location viewpoints can be dynamically synthesized by the edge server for a navigating VR user using geometric signal processing techniques from multi-view imaging such as depth-based view interpolation \cite{ChakareskiVS:12}. Similarly, analysis can be carried out to establish the smallest number of captured \tsd video viewpoints and their spatial distribution over the scene to enable a required quality level for such intermediate interpolated viewpoints.


\subsection{Edge-based operation and mobile edge computing}
Streaming the content from an edge server (bringing the content closest to the user) will minimize the delivery latency compared to traditional server-client network architectures. Moreover, the edge server's computing capabilities can be leveraged to relieve the computing requirements induced upon a mobile VR headset. The former's much more powerful multiple GPUs can be effectively used to this end. In addition to the content decoding computation noted earlier, another computational requirement induced upon a client device running a virtual reality application represents the dynamic rendering of the present field of view of the user (aka viewport). This introduces an additional delay component into the entire end-to-end operational latency chain. Sharing the conventional headset decoding and rendering computation with a nearby edge server can be explored to optimize the delivered immersion fidelity, while meeting the strict system latency of the application. Raw video transmission and scalable 360$^\circ$ tiling can synergistically integrate with this objective, as well.

\subsection{System Design and Integration Embodiment}
\label{sec:SystemDesignArenaIntegrationEmbodiment}
\begin{figure}[htb]
  \centering
  \includegraphics[width=0.65\linewidth]{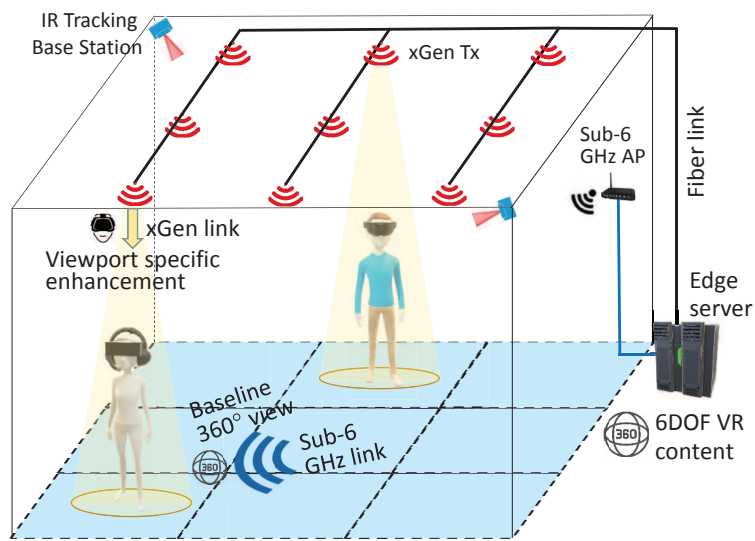}
  \caption{An indoor arena integration of the envisioned virtual reality system. AP: access point; Tx: transmitter}
  \vspace{-0.2cm}
  \label{fig:SystemIntegrationDesign}
\end{figure}

We highlight here in greater detail how a system level integration of the techniques and technologies outlined heretofore can be pursued towards realizing the envisioned VR system concept in practice. To facilitate the discussion, we have included a high level illustration of an arena embodiment of this future application system in Figure~\ref{fig:SystemIntegrationDesign}. There can be multiple traditional sub-6 GHz wireless access points mounted on the walls of the arena and a higher number of emerging high-frequency wireless transmitters mounted on the ceiling of the arena. Both types of transmitters can be linked via fiber optical links to the edge server installed on site. The server can control the dynamic transmission scheduling of the traditional access points and next generation (xGen) transmitters. The user headsets will be equipped with respective sub-6-GHz/xGen dual connectivity receivers. Moreover, the edge server will be equipped with high-end storage and computing capabilities to execute the envisioned dual-connectivity 360$^\circ$ video streaming. In the case of on-demand remote immersion, the compressed 6DOF VR or \tsd video content can already reside at the edge server and be deployed there ahead of time. In the case of live remote immersion, the content can be deployed in real time over a high-speed fiber optic Internet link to the edge server, from the remote location where it would be captured and compressed using scalable \tsd tiling. 

To track the spatial location and head orientation of the users, IR tracking base stations can be mounted on the walls or ceiling of the arena. Present commercially available products of this nature provide accurate positional tracking of less than 1-mm error at temporal resolutions of 250 samples per second or higher. The tracking information is relayed to the edge server in real-time over high-speed links to enable its operation. The spatial area of the arena can be conceptually split into multiple small sectors or cells, and one xGen transmitter can be mounted above each such sector, as illustrated in Figure~\ref{fig:SystemIntegrationDesign}. At the onset of a streaming session for remote immersion, a user can be assigned to a given sector and its xGen transmitter. 
During the session, the user will dynamically navigate the 6DOF VR or \tsd video content and may move across the arena. If this movement is limited within the original sector, the same xGen transmitter can be dynamically steered towards the user, as explained earlier.


On the other hand, as such high-frequency links experience strong signal degradation over short distances, 
dynamically (re)assigning another xGen transmitter to the user can be carried out, every time she will exit her present sector as the session evolves. This can help maximize the received signal quality and thus data rate on the high frequency link of the user, and minimize the interference to other high frequency links in the arena serving other users, at the same time. Thereby, the quality of experience of all simultaneous users in the arena can be augmented. Finally, if more than one user happen to transiently pass through or reside in the same sector at the same time, during the course of the session, the resources of the high-frequency xGen transmitter of that sector can be split across these users during that time, uniformly or preferentially, for some alternative transmitter assignment and steering settings/methods for the system\footnote{We recall that the system design to exclusively serve each user by one xGen transmitter via its narrow transmission beam that is dynamically directed towards the user, as outlined in the Section "High-Frequency Directional Wireless Transmission" earlier, is the default choice that we focus on as the most adequate and performing for the next generation VR system we investigate.}


In particular, we highlight briefly here the prospective options for xGen transmitter steering and (re)assignment in our system, to provide further clarity and instruction to the reader. In the case of FSO wireless technology, electronic, mechanical, or hybrid electro-mechanical transmitter steering can be carried out. (i) The first approach is also known as electronic transmitter switching or assignment, as it effectively activates an FSO transmitter associated with a spatial cell into which a user enters, and deactivates the FSO transmitter associated with the cell that the user exited. This approach comprises the same number of FSO transmitters as spatial cells and their beam width is selected to ensure coverage of the entire spatial area of their respective cells, while minimizing the overlap with transmitter beams of adjacent cell, to minimize the inter-cell interference. Here, the FSO transmitters are fixed (static) in orientation and point directly downwards to the spatial area of the cells. Lastly, when multiple users are present in the same cell, the transmission resources of the respective FSO transmitter can be shared across them using multi-access techniques, e.g., time-division multiple access (TDMA). (ii) Employing mechanical steering requires having at least the same number of transmitters as users in the arena, if there are more users than spatial sectors/cells in the arena\footnote{In the experimental evaluation, we address this choice simply by setting the number of xGen transmitters per cell to be equal to $\lceil N_u/N_c \rceil$, where $N_u$ and $N_c$ denote respectively the number of spatial cells and users in the arena system, and $\lceil \cdot \rceil$ denotes the ceiling operator.}. In this case, there can be multiple transmitters mounted above the spatial area of a single cell and their orientation is dynamically steered towards their respective users to whom they have been assigned using servo motors, as introduced earlier. Periodic reassignment of users to transmitters over the course of a session can be carried out here as well, as the users move across the spatial area of the arena during the course of a media session, to enhance the system performance, as noted a little earlier. (iii) Finally, the hybrid electro-mechanical steering aims to integrate both prior approaches, where a user can be tracked by his or her (re)assigned transmitter using mechanical steering until the user enters a cell where another user is already present. In that case, both users will start to be served using a single static transmitter mounted above the center of that cell using a multi-access technique. Once a user becomes alone again in his or her present cell, the system will switch back to mechanical steering of the respective transmitter assigned to the user, towards the user.

In the case of mm-wave wireless technology, electronic beam formation and steering can be employed to ensure transmitter-receiver alignment as the users move across the arena, as noted earlier in the Section "High-Frequency Directional Wireless Transmission". Here, one option can be to have at least as many transmitters as users in the arena, if there are more users than spatial cells in the arena. Thus, potentially there can be multiple transmitters mounted above each cell in the arena system. Alternatively, the system can comprise the same number of transmitters as cells in the arena, mounted above the center of each respective cell. In this latter case, multiple users present in the same cell can be served by the same transmitter of that cell using multi-beam formation and steering, to share its transmission resources. Multi-beam mm-wave transmitters or access points are prevalently encountered in practice. Finally, in each case, dynamic reassignment of users to transmitters can be carried out during the course of the session as the users move across the spatial area of the arena, to enhance the system performance.


The benefits and shortcomings of the transmitter steering and (re)assignment methods available for each xGen wireless technology and outlined here, in the context of our system, have been explored recently in \cite{ChakareskiKRB:20,ChakareskiKRB:21}.

\section{Example End-to-end Analysis}
We guide the reader here through several high-level analysis examples that build upon each other to highlight the prospective benefits of the envisioned VR system and motivate new studies that can help advance it further. In particular, we first illustrate how the \tsd video tiling and the ability to collect user navigation data can facilitate the development of statistical models of user navigation. In conjunction, we illustrate how effective rate-distortion modeling of the spatiotemporal encoding characteristics of the \tsd panorama across its tiles can be pursued. We then highlight how these modeling advances can be integrated to pursue further analysis that captures the fundamental trade-offs between computing, communication, and signal representation in the context of the end-to-end application performance enabled by the system. Finally, we illustrate how rigorous system level optimization can be built upon these integrated modeling and analysis advances to select key resource allocation decisions.

\begin{figure}[htb]
	\centering
	\begin{subfigure}{0.45\columnwidth}
		\centering
		\includegraphics[width=\linewidth]{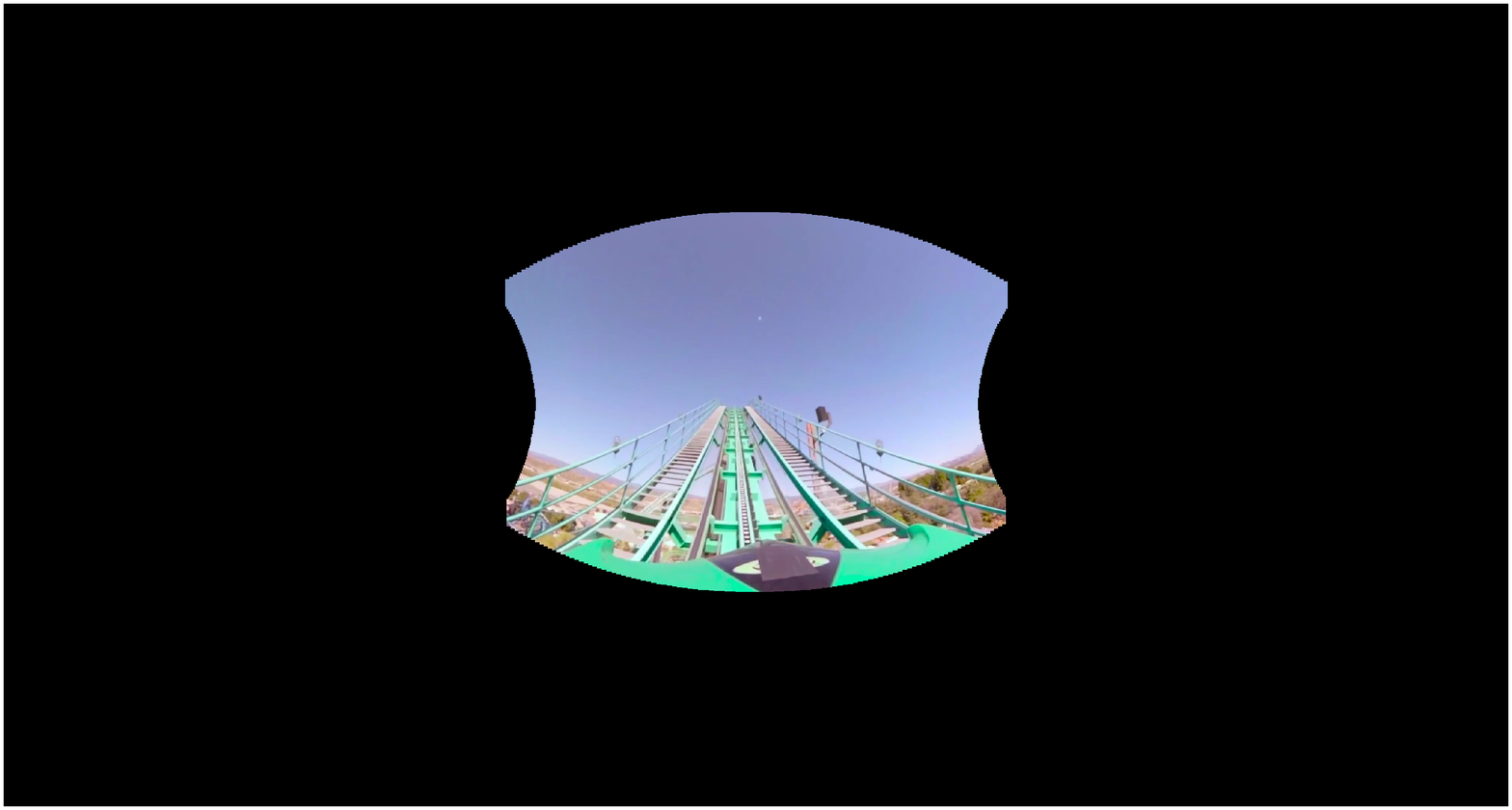}
        $(\varphi,\theta) = (0^{\circ},0^{\circ})$
	\end{subfigure}
	\begin{subfigure}{0.45\columnwidth}
		\centering
		\includegraphics[width=\linewidth]{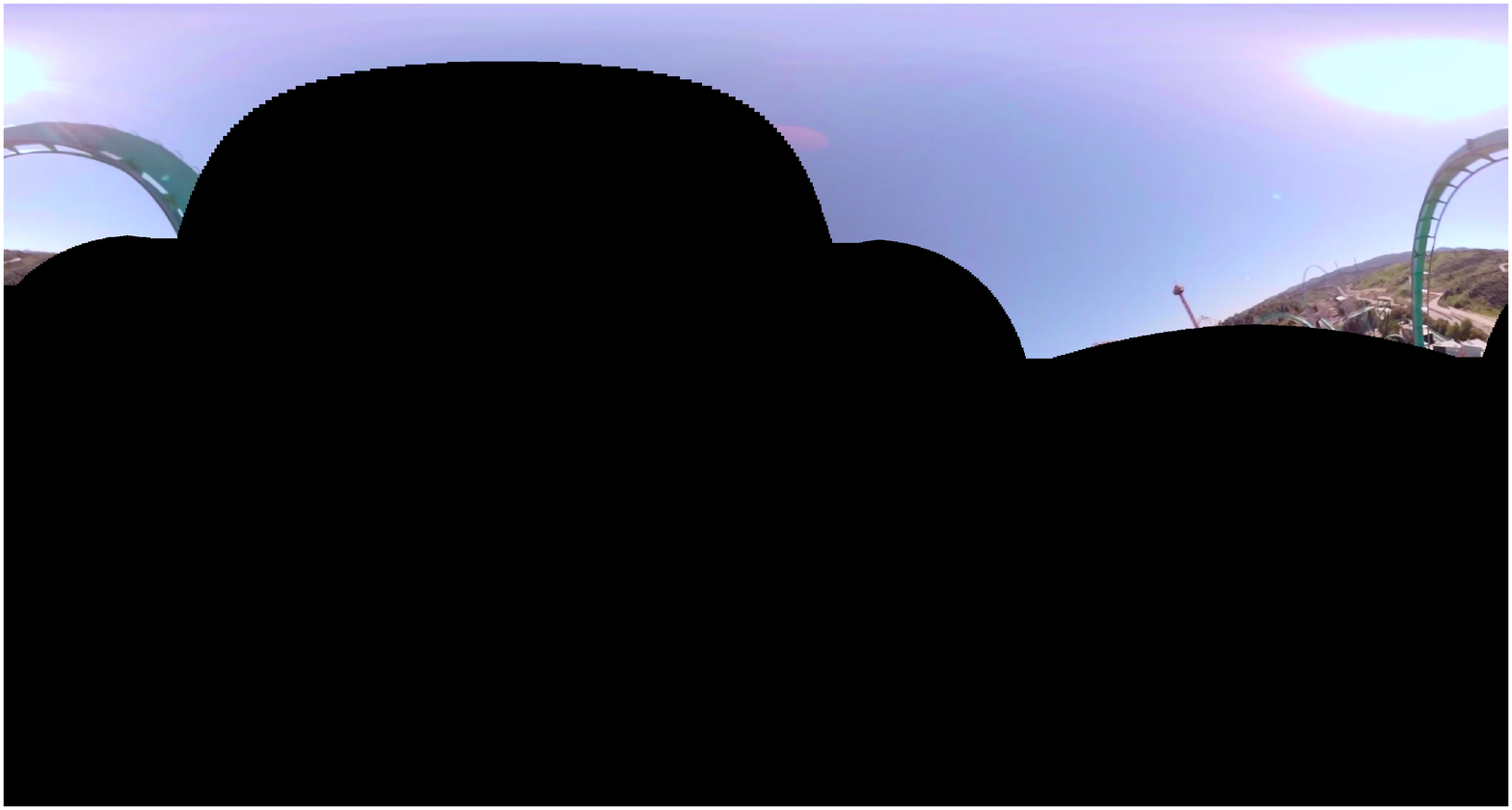}
        $(\varphi,\theta) = (120^{\circ},-60^{\circ})$.
	\end{subfigure}
    \caption{The viewport's latitude coordinate on the \tsd view sphere (polar angle $\theta$ = head rotation angle pitch) impacts its shape and size on the \tsd panorama.}
    \label{fig:ViewportSizeLatitudeImpact}
\end{figure}

{\bf Navigation modeling.} The head movement navigation actions of a user can be collected over time at the edge server and can be used to characterize the probability that a \tsd tile will appear in the user viewport during a time period. This will inform the understanding of how important each tile and its content are for the quality of experience of the user. One way to capture this information is to integrate the number of instances a tile will appear in the viewport during navigation of the content over the time interval of interest. This approach can be made rigorous by identifying the spatial portion $S^{nm,V_c}_j$ of each \tsd tile $(n,m)$ present in the user viewport $V_c$ at the temporal display instance $t_j$ of the $j$-th panoramic video frame comprising the \tsd content. Care must be taken here to account for the widely varying viewport size depending on its latitude on the \tsd panorama, to integrate this information properly then over time. This phenomenon is induced by the equirectangular mapping of the native spherical data that is employed at compression, as explained earlier. It is illustrated in Figure~\ref{fig:ViewportSizeLatitudeImpact} that highlights the shape and size of a viewport on the \tsd panorama for two different yaw and pitch rotation angles, denoted therein as azimuth and polar angles $(\varphi,\theta)$ in spherical coordinates.

Formally, the quantity $S^{nm,V_c}_j$ represents the overlap or intersection between the spatial areas of the tile and the viewport in the \tsd panorama at that time instance and can be captured as $S^{nm,V_c}_j = S^{V_c}_j \cap S^{nm}_j$, where $S^{nm}_j$ and $S^{V_c}_j$ denote the sets of pixels representing the two denoted spatial areas. To account for the unequal viewport size on the equirectangular plane across time, in developing a statistical model of user navigation, one can normalize the fractions of the spatial areas of every tile present in the user viewport $V_c$ at $t_j$, using $s^{nm}_j = \frac{|S^{nm,V_c}_j|}{\sum_{n,m}|S^{nm,V_c}_j|}$,
\noindent where $|S|$ denotes the size of a set $S$, in this case in number of pixels. Thus, $\{s^{nm}_j\}$ represents the normalized distribution of the spatial area of the user viewport across every tile in the $360^\circ$ panorama, at time instance $t_j$.

Finally, given $\{s^{nm}_j\}$, one can formulate the probability (likelihood) of the user navigating tile $(n,m)$ over a time interval spanned by the time instances $[t_i,t_j]$, as $P_{nm}^{(t_i,t_j)} = \frac{\sum_{k=i}^{j}s^{nm}_k}{j-i+1}$.
In other words, $P_{nm}^{(t_i,t_j)}$ indicates how likely tile $(n,m)$ will appear (at least in part) in the user viewport during navigation of the $360^\circ$ video from its temporal instance $t_i$ to its temporal instance $t_j$, or the popularity of the $360^\circ$ scene content captured by the tile for these user and time interval. For instance, if $t_i$ and $t_j$ correspond to the first and last video frame of the $360^\circ$ video, then, $P_{nm}^{(t_i,t_j)}$ captures the navigation probability or popularity of tile $(n,m)$ across the entire video.

\begin{figure}[htb]
\centering
\begin{subfigure}{.5\columnwidth}
  \centering
  \includegraphics[width=1\linewidth]{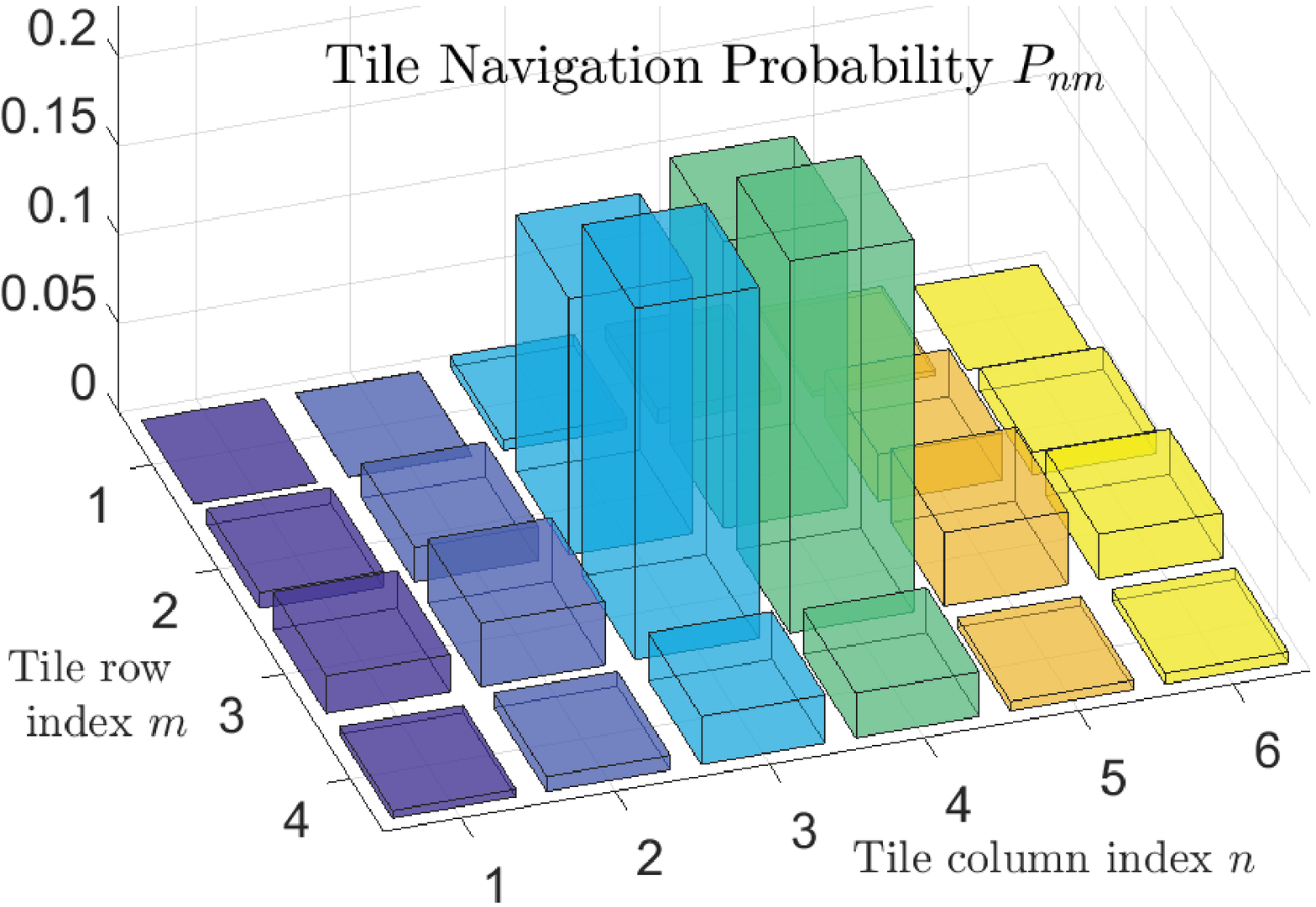}
\end{subfigure}%
\begin{subfigure}{.5\columnwidth}
  \centering
  \includegraphics[width=1\linewidth]{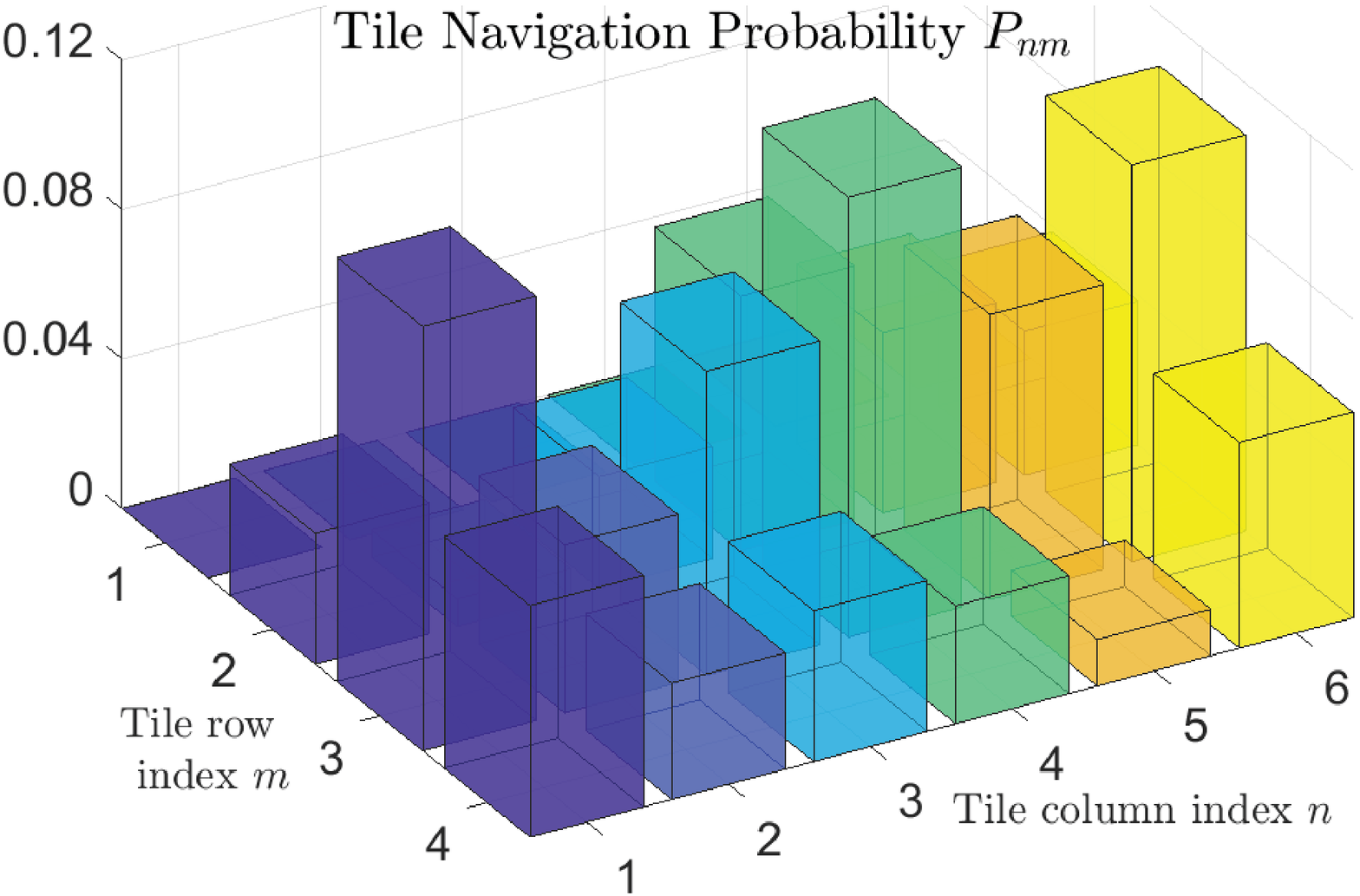}
\end{subfigure}
\vspace{-0.3cm}
\caption{Expected $\{P_{nm}\}$ for $360^\circ$ tiles for Roller Coaster (left) and Wingsuit (right).}
\label{fig:RollerCoasterWingsuitNavigationProbabilities}
\vspace{-0.3cm}
\end{figure}

To highlight the nature of these quantities, we illustrate in Figure~\ref{fig:RollerCoasterWingsuitNavigationProbabilities} their expected values (averages across a large user set) for two popular \tsd videos used in our analysis, where we considered a popular $6 \times 4$ tiling\footnote{Denser tiling layouts increase the processing complexity and reduce the compression efficiency, but enable more precise delineation of the user viewport and thus more accurate analysis and resource allocation.}. Concretely, one can observe that in the case of Roller Coaster, video tiles on the fringes of the $360^\circ$ panorama are rarely navigated by a user, i.e., they scarcely appear in the user's viewport, during a $360^\circ$ video navigation session. Conversely, video tiles indexed as $(n,m) =$ (3,2), (4,2), (3,3), and (4,3) are quite often navigated by the user, as noted by their much higher navigation likelihoods shown in the histogram.

The navigation probabilities of $360^\circ$ video tiles for the video Wingsuit, induced by users experiencing the content, look fairly different. This is due to the nature of this content and the induced specific interests of the users, expressed when navigating it. In particular, one can see here that the typical user is predominantly interested in navigating the southern hemisphere of the $360^\circ$ panorama, as seen from the respective tile navigation probabilities indicated in the histogram. The more dynamic and interesting content of Wingsuit resides spatially in the southern hemisphere of its $360^\circ$ view sphere.

To address the prospect of imperfect or lack of knowledge of near future navigation actions and introduce further robustness into the system, e.g., in contexts such as live remote immersion, predicting the upcoming head movement (or jointly head and body movement) navigation actions or 360-degree tile navigation likelihoods for a user can be integrated. This can be carried out accurately via regression or machine learning techniques, based on the user’s navigation history and prospectively including features extracted from the already navigated content \cite{QianHJG:16,HouDZB:21}. It has been shown that the navigation actions of a user exhibit a strong low-frequency component and short-term correlation over time that can benefit such prediction methods. Moreover, it has been observed that the dynamics of the navigation actions of a user generally correlate well with the temporal characteristics of the content that is navigated \cite{BlandinoRCCKG:21}.

\begin{figure}[htb]
\centering
\begin{subfigure}{.5\columnwidth}
  \centering
  \includegraphics[width=1\linewidth]{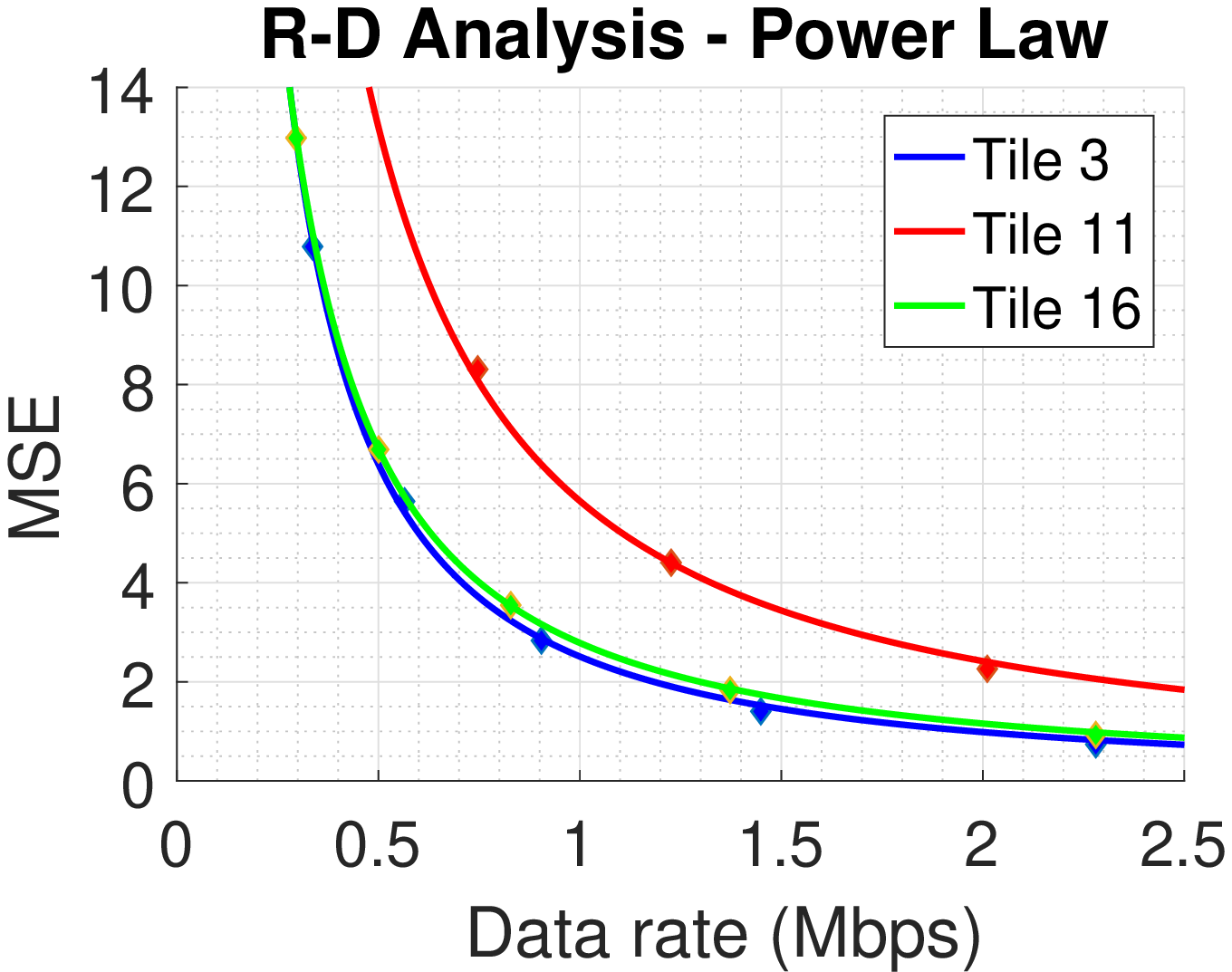}
\end{subfigure}%
\begin{subfigure}{.5\columnwidth}
  \centering
  \includegraphics[width=1\linewidth]{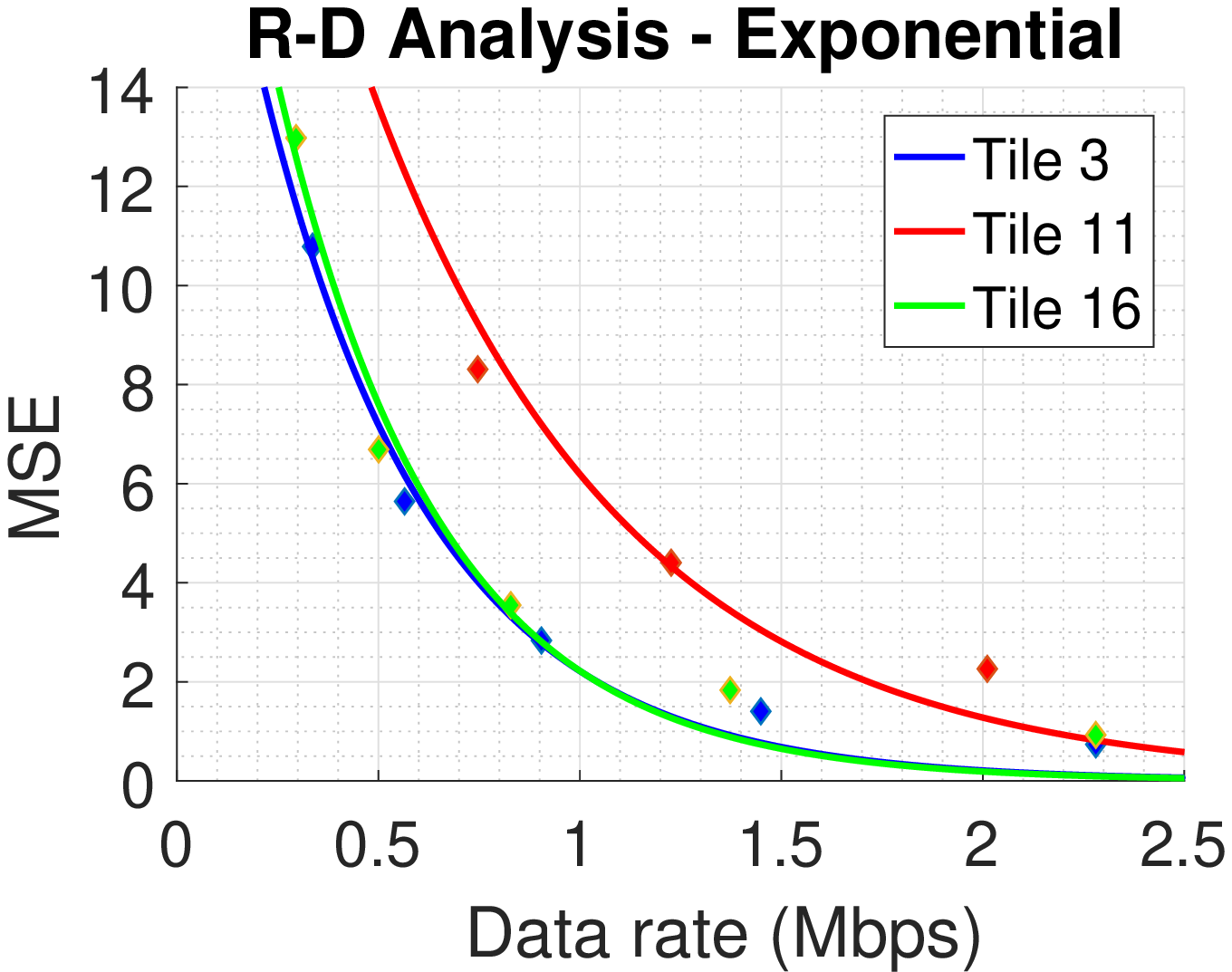}
\end{subfigure}
\vspace{-0.3cm}
\caption{Rate-distortion dependency $D(R)$: (Left) Power law model and (Right) Exponential model. Actual data points shown as markers.}
\label{fig:TileRateDistortionDependency}
\vspace{-0.3cm}
\end{figure}

{\bf Rate-distortion modeling.} One can also accurately model the rate-distortion dependency of compressed GOP tiles to facilitate analysis and optimization of the edge server's operation. Here, we examine two models, exponential ($D = c_1 e^{-d_1 R}$) and power law ($D = c_2 R^{d_2}$), to capture this dependency. Concretely, we vary the encoding rate $R$ of a tile and record its respective Mean Squared Error (MSE) reconstruction distortion $D$. We graph the obtained pairs $(R,D)$ in Figure~\ref{fig:TileRateDistortionDependency} with markers, for three representative tiles of the popular 360$^\circ$ video Roller Coaster. The two analytical models are also graphed in Figure~\ref{fig:TileRateDistortionDependency}. We can see that Tiles 3 and 16 have steeper rate-distortion dependency, due to their relatively static content. Tile 11 requires higher encoding rate to achieve the same reconstruction error, due to its more dynamic content. Figure~\ref{fig:TileRateDistortionDependency} shows that the power law model provides a more accurate characterization across all three tiles. This motivates its use in our later analysis.

We note that in video compression the raw pixel data is stored in the three-dimensional luminance-chrominance color space denoted as YUV, to enable data volume reduction. The Y (luminance) component captures exclusively the signal strength of each pixel, i.e., its monochromatic intensity or brightness. The two U and V chrominance components capture exclusively the color information associated with each pixel. Video signal distortion or quality is commonly measured only on the Y component due to its nature.

{\bf Motivating resource allocation example.}
Building upon the two modeling advances and scalable \tsd tiling, we illustrate here how data rate resources can be effectively allocated over the \tsd panorama and time. When we examine and formulate the end-to-end performance of the envisioned VR system concept later, we will integrate this motivating example analysis therein.

When the \tsd content is compressed using scalable tiling, the data rate and reconstruction error of a GOP tile $(n,m)$ can be directly related to the number of embedded layers $l_{nm}$ selected to represent the tile. These relationships can be made precise using the rate-distortion modeling described earlier. Similarly, given the statical navigation profile $\{P_{nm}\}$ of a user for this GOP, one can formulate the expected viewport distortion experienced by the user during the GOP as $\sum_{nm}P_{nm}D_{nm}(l_{nm})$. Finally, given a maximum streaming data rate of $C$, one can then aim to optimally select the number of layers $l_{nm}$ sent for every tile $(n,m)$ during the GOP time interval, such that the expected user viewport distortion over that interval is minimized. This can be formally captured via the optimization:
\beq
\min_{\{l_{nm}\}}\sum_{nm}P_{nm}D_{nm}(l_{nm}), \, \text{subject to:} \sum_{nm}R_{nm}(l_{nm})\le C. \nonumber
\eeq

%
%

\begin{figure}[htb]
\centering
\includegraphics[width=1.1\figwidth]{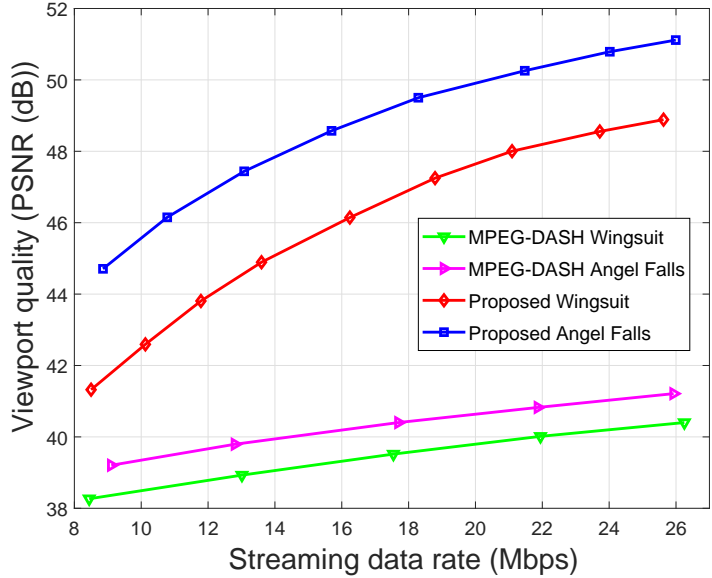}
\vspace{-0.4cm}
\caption{$360^\circ$ video transmission efficiency of the motivating resource allocation example vs. MPEG-DASH.}
\label{fig:TransmissionEfficiencyVsMPEG-DASH}
\end{figure}

To highlight its benefits and potential, we implemented the optimization above, varied $C$, and computed the optimal solution and the respective minimum expected viewport distortion in each case (the value of the objective function at the optimal solution). Simultaneously, we recorded in the same context the corresponding performance of the state-of-the-art MPEG-DASH streaming standard, implemented with spatial viewport-adaptation \cite{Petrangeli2017}. We examined two popular 4K $360^\circ$ videos Wingsuit and Angel Falls to carry out this motivational assessment. We show these outcomes in Figure~\ref{fig:TransmissionEfficiencyVsMPEG-DASH} where the y-axis indicates the delivered viewport quality or immersion fidelity via a base-10 logarithmic inverse of the expected distortion, formally known as the {\em peak-signal-to-noise ratio} (PSNR). We note that throughout this paper, the reported PSNR results are measured with respect to the luminance (Y) component of the user viewport video signal, as explained earlier. On rare occasions, these results are denoted as Y-PSNR to indicate this aspect.

We can see that the approach highlighted above enables considerable benefits over MPEG-DASH by integrating the user navigation actions and the rate-distortion trade-offs across the $360^\circ$ panorama, in deciding how transmission resources should be allocated. Approximately 6-7 dB of immersion fidelity improvement have been enabled across both $360^\circ$ videos and all network bandwidth values considered in Figure~\ref{fig:TransmissionEfficiencyVsMPEG-DASH}. These advances can in turn enable much higher operational efficiency for a streaming system for $360^\circ$ video delivery that integrates this methodology, as our subsequent analysis will highlight.

The problem formulation can be easily extended to integrate tile-level weights $w_{nm}$ that will capture the amount of deformation (stretching) that the content of each tile $(n,m)$ undergoes on the \tsd sphere when mapped back for viewing on the VR headset. This deformation is solely dependent on the tile's latitude and emphasizes equatorial tiles as more sensitive to content distortion because of this. By this integration, the formulation will capture as its objective the expected tile-level weighted to spherically-uniform viewport distortion, whose pixel-level deterministic counterpart denoted as WS-MSE has recently been introduced as a more adequate quality metric for omnidirectional content \cite{SunLY:17}.





{\bf Per-user end-to-end performance analysis.} The edge server in the envisioned VR system can benefit from a navigation profile $\{P_{nm}\}$ for a GOP to preferentially treat the tiles comprising it when streaming the content to the respective user, as motivated by previous resource allocation example. Concretely, the server can first identify as $M$ the set of GOP tiles with non-zero navigation likelihoods. To construct the baseline content layer, the server can assign streaming data rates $R_{nm,w}$ to lower indexed embedded layers from the scalable \tsd tiling exclusively for tiles $(n,m)$ from $M$, to maximize the resource utilization.
If $\Delta T$ is the temporal duration of a GOP, then the latency of streaming the baseline content layer to a user over her traditional wireless connectivity link can be formulated as $\tau^w = \frac{\sum_{(n,m)\in M}R_{nm,w}\Delta T}{C^w}$, where $C^w$ indicates the transmission capacity of the link.

Similarly, by facilitating $\{P_{nm}\}$, the server can identify a subset $M^{\text{r}}$ of tiles in $M$ that are most relevant for the quality of experience of the user. The construction of $M^{\text{r}}$ can be made rigorous as explained a little later. These tiles can be streamed as raw data as part of the enhancement content layer, to maximize their impact. They can be reconstructed first as such at the edge server at the highest quality, from the compressed scalable \tsd tiling that will reside there, by decoding each compressed tile from $M^{\text{r}}$ from all its embedded layers. If the encoding data rate of the entire set of layers for tile $(n,m)$ is $R_{nm,\max}$, then the decoding latency of reconstructing tiles from $M^{\text{r}}$ as raw data at the server can be formulated as $\tau^Z = \frac{\sum_{(n,m)\in M^{\text{r}}}R_{nm,\max}\Delta T}{Z}$, where $Z$ is the decoding speed of the server.

The construction of the enhancement content layer by the server can be completed by selecting an adequate set of higher indexed embedded layers from the scalable \tsd tiling for the remaining (compressed) tiles in $M$ (the set $M^{\text{e}} = M \setminus M^{\text{r}}$). Let $R_{nm,x}$ be the selected streaming data rate for each such tile in the enhancement content layer.
This content layer will be streamed to the user over her xGen wireless connectivity link characterized with transmission capacity $C^x$. The associated transmission delay can be characterized as $\tau^x = \frac{|M^{\text{r}}|E_r + \sum_{(n,m)\in M^{\text{e}}}R_{nm,x}\Delta T}{C^x}$, where $E_r$ denotes the data size of a raw GOP tile.


Finally, the latency on the client device involves decoding compressed tiles and rendering the viewport. These tasks will be carried out twice for the baseline and enhancement content layers streamed to the user, to enable respectively application reliability and high quality immersion, as introduced earlier. The decoding capability (speed) $z$ of the headset will need to be partitioned between decoding the baseline content layer and decoding the subset of compressed tiles from the enhancement content layer $M^{\text{e}}$. Let these two decoding speed allocations be denoted as $z^w$ and $z^x$, respectively. Thus, the required time to carry out each of these two decoding tasks can be formulated as $\tau^{z,w} = \frac{\sum_{(n,m)\in M}R_{nm,w}\Delta T}{z^w}$ and $\tau^{z,x} = \frac{\sum_{(n,m)\in M^{\text{e}}}R_{nm,x}\Delta T}{z^x}$. Similarly, the processing capability (power) $r$ of the headset will need to be partitioned between rendering the viewport solely from the baseline content layer and rendering the viewport at enhanced quality jointly from the baseline and enhancement content layers. Let these two processing power allocations be denoted as $r^w$ and $r^x$, respectively. The time delay of carrying out each of these two rendering tasks can be formulated as $\tau^{r,w}  = \frac{E_v}{r^w b_h}$ and $\tau^{r,x}  = \frac{E_v}{r^x b_h}$, where $E_v$ is the data size of the viewport after decoding and $b_h$ is the computed data volume per CPU cycle on the headset.

Benefiting from the earlier modeling, the quality of immersion delivered to a VR user by the envisioned dual-connectivity system can be accurately captured by formulating the expected viewport distortion experienced by the user as:
$$D\left(\{R_{nm,w},R_{nm,x}\}\right) = \sum_{(n,m)\in M^{\text{r}}} P_{nm} a_{nm} R_{nm,\max}^{b_{nm}} + \sum_{(n,m)\in M^{\text{e}}} P_{nm} a_{nm} \left(R_{nm,x} + R_{nm,w}\right)^{b_{nm}},$$
where $a_{nm}$ and $b_{nm}$ are the rate-distortion power law model parameters for tile $(n,m)$. Additionally, each summation term above can be weighted by a tile-level WS-MSE coefficient $w_{nm}$ that captures the unequal impact of each tile $(n,m)$ when the panoramic content is re-projected back on the \tsd sphere for viewing, as introduced earlier. 

The analysis of latency and quality of immersion can enable pursuing end-to-end optimization of the system. One way of formally capturing this is via the problem formulation highlighted below that aims to maximize the quality of immersion delivered to a user given various system latency and resource constraints. To make the notation more compact, we henceforth replace the symbols $(n,m)$ with $(i,j)$.
\begin{align}
& \mathop{\min_{\{R_{ij,w}\},\{R_{ij,x}\},}}_{M^{\text{r}},\{z^w,z^x\},\{r^w,r^x\}} D\left(\{R_{ij,w},R_{ij,x}\}\right), \label{eq:pf1}\\
\hspace{-0.3cm} \text{subject to:} \quad & \tau^w + \tau^{z,w} + \tau^{r,w} \le \Delta T, \label{eq:c1}\\
& \tau^Z  + \tau^x + \tau^{z,x} + \tau^{r,x} \le \Delta T, \label{eq:c2}\\
& R_{ij,w} \in [R_{ij,\min},R_{ij,\max}], \; R_{ij,x} \le R_{ij,\max} - R_{ij,w}, \label{eq:c3}\\
& r^w + r^x \le r, \quad z^w + z^x \le z. \label{eq:c5}
\end{align}

The inequalities \eqref{eq:c1} and \eqref{eq:c2} capture the application latency requirements associated with the transmissions on the two wireless connectivities. The inequalities in \eqref{eq:c3} capture the data rate limits enabled by the scalable \tsd tiling, where $R_{ij,\min}$ is the smallest possible data rate for tile $(i,j)$ (the data rate of its first embedded layer). We note that the transmission latency constraints \eqref{eq:c2} and \eqref{eq:c3} are stricter than and imply the respective transmission capacity constraints on the two wireless connectivity links. Thus, the latter two constraints do not need to be included in the optimization (\ref{eq:pf1})-\eqref{eq:c5}. Finally, the inequalities in \eqref{eq:c5} capture the limited computational resources of the VR headset of the user.

The problem (\ref{eq:pf1})-\eqref{eq:c5} is mixed-integer programming, which is hard to solve optimally in practice. Still, with a clever selection of $M^{\text{r}}$, one can pursue the optimal solution at lower complexity. In particular, for a given set of tiles to be sent as raw data, the problem above can be transformed to geometric programming and solved exactly via iterative approaches that converge rapidly \cite{xu2014global}. Moreover, though the number of tiles in $M$ is not excessive, instead of pursuing a combinatorial approach to identify the optimal $M^{\text{r}}$ as part of the overall solution to (\ref{eq:pf1})-\eqref{eq:c5}, one can accomplish that more effectively by sorting the tiles in $M$ according to an adequate criterion, related to the objective function, and constructing $M^{\text{r}}$ to comprise the first $k$ tiles from the sorted $M$. By sweeping $k$ across a range of values, solving (\ref{eq:pf1})-\eqref{eq:c5} for each such thereby constructed $M^{\text{r}}$ using geometric programming, and identifying the optimal value of the objective function in each case, one can then identify the overall solution.

Moreover, though $k$ can technically range up to $|M|$, the maximum number of tiles that can be sent as raw data will typically be smaller, as otherwise the system latency constraints will be violated. This is a result of the large data size of a raw GOP tile and the present transmission capacities of xGen technologies. Yet, when the latter are selected rather too conservatively in (\ref{eq:pf1})-\eqref{eq:c5}, even sending one raw tile can become unfeasible, and the choice of $k = 0$ lends itself as optimal. Finally, we observed that selecting the navigation likelihood weighted derivative of the rate-distortion dependency for a GOP tile as the criterion for sorting the tiles in $M$ in decreasing order leads to the same overall solution produced when instead a combinatorial search is used to identify $M^{\text{r}}$.

The optimization described in this section would be applied at the edge server to every subsequent GOP of the content transmitted to a given user. In a subsequent section, we will build upon the analysis and optimization developed here, to formulate an overall multi-user optimization strategy for the envisioned system.

{\bf Dynamic xGEn transmitter assignment.} Earlier, as part of the arena embodiment of the envisioned system outlined in Section~\ref{sec:SystemDesignArenaIntegrationEmbodiment}, we highlighted a prospective dynamic reassignment of xGen transmitters to mobile VR users during the course of a session for further performance enhancement, to complement the multi-user system level resource allocation optimization described next. In particular, during a session, as the mobile VR users navigate the content across the spatial area of the arena, such reassignment can be carried periodically such that, for instance, the smallest signal-to-noise ratio (SNR) across all transmitter and user pairs in the arena, is maximized. As SNR is inversely proportional to the distance from a user and an xGen transmitter, an equivalent approach can be pursued aiming to minimize the longest distance between a user and a transmitter, in an assignment. Either of these approaches will lead to a combinatorial optimization problem that may be complex to solve, depending on the number of users and transmitters in the system.

An alternative lower-complexity strategy would be to formalize the assignment as a graph-bottleneck matching on a bipartite graph comprising the users and xGen transmitters, as the two vertex set partitions, and where the graph edge weights would correspond to the distances between the users and transmitters. Concretely, in this case one will seek to find the maximum size vertex matching on the graph, whose biggest edge weight is smallest across all the matchings of that size. An efficient algorithmic solution can then be formulated that will integrate and benefit from the Dulmage–Mendelsohn decomposition of such a graph, to construct the optimal assignment $\pi$ incrementally. Our recent advances demonstrate the benefits of this strategy \cite{ChakareskiK:21}. Moreover, we have empirically observed that for the diverse settings considered in \cite{ChakareskiK:21} using the {\em max min SNR} as an objective metric for the transmitter assignment enables virtually the same performance, at much lower computational cost for the optimization, over the case of using the {\em max min SINR} (Signal-to-Interference and Noise Ratio). The performance evaluation that we carry out subsequently integrates the impact of interference via the SINR.

%
{\bf Multi-user system performance analysis and optimization.} The overall objective of the system is to maximize the delivered immersion quality across all users in the arena. Once users are assigned to xGen transmitters, as described above, this objective can be pursued by formulating and solving a respective optimization problem of interest. Building upon the notation and formalism introduced earlier in Section "Per-user end-to-end performance analysis", this multi-user problem can be formally described as:

\begin{align}
\text{(MU-OPT):} \quad & \min_{\left\{ \substack{\{R_{ij,w}^u\},\{R_{ij,x}^u\}, C_u^w, C_u^x, \\ M^{\text{r}}_u, \{z_u^w,z_u^x\},\{r_u^w,r_u^x\}, Z_u } \right\}_{\forall u}} \sum_u D_u\left(\{R_{ij,w}^u,R_{ij,x}^u\}\right), \label{eq:multi-user-optim}\\
\hspace{-0.3cm} \text{subject to:} \quad & \eqref{eq:c1} - \eqref{eq:c5}, \nonumber\\
& \sum_u Z_u \le Z, \label{eq:AggregateComputingCapacity}\\
& \sum_u C_u^w \le C^w, \label{eq:AggregateTransmissionCapacityLowFrequency}\\
& \sum_{u \in U_t} C_u^x \le C^{x,t}, \forall t, 
\label{eq:AggregateTransmissionCapacityHighFrequencyPerTransmitter}
\end{align}

\noindent where we have introduced a subscript or superscript $u$ to all symbols used in (\ref{eq:pf1})-\eqref{eq:c5} earlier, to indicate a specific user $u$ here. Moreover, $Z$ denotes the aggregate decoding capacity of the edge server in \eqref{eq:AggregateComputingCapacity}, $C^w$ denotes the aggregate (down-link) transmission capacity of the traditional wireless technology in \eqref{eq:AggregateTransmissionCapacityLowFrequency}, and $C^{x,t}$ denotes the aggregate (down-link) transmission capacity of the xGen transmitter $t$ in \eqref{eq:AggregateTransmissionCapacityHighFrequencyPerTransmitter}. Finally, alike in \eqref{eq:AggregateTransmissionCapacityHighFrequencyPerTransmitter}, $U_t$ denotes the set of users assigned to xGen transmitter $t$.

Solving the optimization problem above (MU-OPT) in its full generality can be pursued via iterative techniques and the Lagrange multiplier method that would build upon the respective solution of the per-user optimization (\ref{eq:pf1})-\eqref{eq:c5} as a key inner step of such approaches. Moreover, solving for the respective Lagrange multipliers associated with the constraints \eqref{eq:AggregateComputingCapacity}-\eqref{eq:AggregateTransmissionCapacityHighFrequencyPerTransmitter} as part of the overall optimization can be carried out using for instance integrated fast sub-gradient techniques. Next, we outline the impact of parameter selection, user-transmitter assignment setting, and the indoor nature of our system on (MU-OPT).

With judicious analysis and choice of key system resource parameters captured by the constraints \eqref{eq:AggregateComputingCapacity}-\eqref{eq:AggregateTransmissionCapacityHighFrequencyPerTransmitter}, the general optimization outlined above can be simplified. Concretely, it is expected that in our system setting the edge server will be equipped with computing capabilities that exceed those of the mobile clients by orders of magnitude and that it will be well provisioned to assist abundantly the expected number of users in the arena system with such capabilities. Thus, one simple approach of dispensing with \eqref{eq:AggregateComputingCapacity} without preventing the integrated per-user optimization (\ref{eq:pf1})-\eqref{eq:c5} from exploring the full range of performance trade-offs and benefits is to set $Z_u = Z/N_u, \forall u$, where $N_u$ is the number of users in the arena. We have empirically verified this assessment in our extensive evaluations. It is rational to follow the same approach to dispense with \eqref{eq:AggregateComputingCapacity} by selecting $C_u^w = C^w/N_u$, with virtually no impact on the optimal performance delivered to each user for the following reasons. First, it is similarly expected that in our system setting the traditional wireless connectivity will be well provisioned to provide plentiful down-link transmission capacity to the expected number of mobile VR users to be served. Second, the envisioned arena system concept has an indoor nature and a relatively small spatial footprint. Third, the traditional wireless connectivity is integrated solely to augment the system/application reliability. Optimizing the choice of $C_u^w$ will not make an impact on this objective.

Finally, dispensing with \eqref{eq:AggregateTransmissionCapacityHighFrequencyPerTransmitter} and addressing the choice of $C_u^x, \forall u,$ is more subtle. On the one hand, in an embodiment setting of the envisioned system that integrates prospective multi-user assignment of its xGen transmitters, as discussed earlier in the last segment of Section "System design and integration embodiment", it is alike expected that each xGen transmitter $t$ will be equipped with ample additional transmission capacity to support the prospective serving of multiple users in parallel and provisioned adequately for the expected number of users to be assigned to a single xGen transmitter during a session. Then, dispensing with the constraint \eqref{eq:AggregateTransmissionCapacityHighFrequencyPerTransmitter} by splitting the transmission capacity of each transmitter $t$ uniformly across its assigned users, i.e., $C_u^x = C^{x,t}/|U_t|$, can be a sensible choice in this context as well, which may not impair notably the truly optimal high-quality immersion that can be enabled for every user in the arena via the assigned xGen transmitter. On the other hand, the rate-distortion dependency $D_u\left(\{\cdot,R_{ij,x}^u\}\right)$ that captures the delivered viewport quality may exhibit different performance trade-offs across different users $u$ over time, depending on the navigation characteristics of the users and the 6DOF content that is explored. This factor may lead to some performance loss depending on the magnitude of the difference between the uniformly assigned capacity $C_u^x$ and its rigorously selected value as an integral part of the optimization with the constraint \eqref{eq:AggregateTransmissionCapacityHighFrequencyPerTransmitter} kept in place. Still, it should be noted that for the present values of the transmission capacity $C_u^x$ that can be assigned to a user and rate-distortion dependencies of existing 3DOF/6DOF \tsd video/VR content, the operational points of a streaming system typically lie in the (far right) saturation portion of the rate-distortion performance dependency. This means that marginal differences in the value of $C_u^x$ assigned to user $u$ will only make a negligible difference in the overall system streaming performance and quality of experience delivered to the user. All the above considerations outlined here related to the choice of $C_u^x, \forall u,$ with the aim to prospectively omit \eqref{eq:AggregateTransmissionCapacityHighFrequencyPerTransmitter} from the optimization need to be carefully considered.

Yet, we have empirically observed in our extensive evaluations that the system setting comprising dynamic steering of the xGen transmitters and an exclusive assignment of one user per transmitter, i.e., the number of users and xGen transmitters in the arena is the same, provides the best performance over other prospective settings of user-transmitter assignment and transmitter steering, outlined earlier in the last segment of Section "System design and integration embodiment". For this characteristic setting of our system, the constraint \eqref{eq:AggregateTransmissionCapacityHighFrequencyPerTransmitter} will be met by default, by simply setting $C_u^x = C^{x,t}$ for the unique user $u$ assigned to that xGen transmitter $t$. This will dispense with the need for \eqref{eq:AggregateTransmissionCapacityHighFrequencyPerTransmitter} and will simultaneously enable delivering the highest possible immersion quality to user $u$. In turn, the latter will also signify that the best overall performance of the system is achieved, since the multi-user optimization (MU-OPT) formulated herein will decouple into multiple independent instances of the per-user optimization (\ref{eq:pf1})-\eqref{eq:c5} applied to every user-transmitter pair $(u,t)$, given the additive nature of the objective function in \eqref{eq:multi-user-optim} and the rationale for dispensing with the other constraints \eqref{eq:AggregateComputingCapacity}-\eqref{eq:AggregateTransmissionCapacityLowFrequency} described earlier. Lastly, beyond the argument carried out regarding the system-level constraints of (MU-OPT), as it applies to our specific context, we note that computational complexity reduction for an optimization method and user fairness are two commonly invoked reasons to justify uniform resource sharing across multiple entities in resource allocation problems, which that can be applicable here too.

Let $\{\text{GOP}_{u,k}\}_{\forall u}$ denote the collection of GOP content that needs to be transmitted to every user $u$ in the system at the onset of the GOP temporal interval $k$ of the 6-DOF remote immersion session. The optimization strategy described here would be applied at the edge server to every subsequent $\{\text{GOP}_{u,k}\}_{\forall u}$, i.e., for $k=1,2,\dots$

{\bf Performance analysis examples.} We highlight some of the key performance trade-offs arising in the arena embodiment setting of the envisioned system by implementing in simulation its most critical components and the optimization (\ref{eq:pf1})-\eqref{eq:c5}, together with the graph matching based strategy for dynamic user to xGen transmitter assignment outlined above. We considered that the spatial area of the arena is 6m $\times$ 4m and is split into six sectors. We further considered that there are six users in the arena and six xGen transmitters mounted on the ceiling above the center point of each cell, as illustrated in Figure~\ref{fig:SystemIntegrationDesign}. The beams of the transmitters are steered towards their assigned mobile users either electronically, in the case of mm-wave wireless technology, or mechanically, in the case of FSO wireless technology, as introduced earlier in the Section "High-frequency directional wireless transmission". We integrated tile-level spherical distortion coefficients $w_{nm}$ into the objective function in (\ref{eq:pf1}), when solving the optimization, and computed the PSNR of the optimal value of the objective function, to examine in our analysis. Henceforth, we denote this performance metric as WS-PSNR to distinguish it from the traditional PSNR, computed from the optimal value of the objective function without the integration of $\{w_{nm}\}$ into the optimization.
To carry out the evaluation, we used the popular 3-DOF 8K 60-fps \tsd video Runner and 6DOF VR content Museum, as well as actual 3-/6-DOF user navigation traces that we collected in our wireless VR arena lab \cite{ChakareskiK:21,ChakareskiASZ:21}. 
Key system performance parameters were selected in the evaluation to correspond to the computational and transmission capabilities, respectively, of present state-of-the-art GPUs, VR headsets, and traditional/emerging wireless technologies 
\cite{CuervoCK:18}. The considered values for these parameters are captured by the ranges of values representing the axes associated with the independent variables in the performance graphs included here.

The first two sets of performance analysis examples consider the \tsd video content. The third analysis example set considers the 6DOF VR content represented as a collection of spatial \tsd video viewpoints that users can dynamically select to navigate as they move across the arena (see Figure~\ref{fig:6DOF_360VideoStreaming}). Each such omnidirectional viewpoint is compressed into scalable \tsd tiling, at 120fps frame rate and 12K spatial resolution panorama.

\begin{figure}[htb]
\centering
\begin{subfigure}{.5\columnwidth}
  \centering
  \includegraphics[width=1\linewidth]{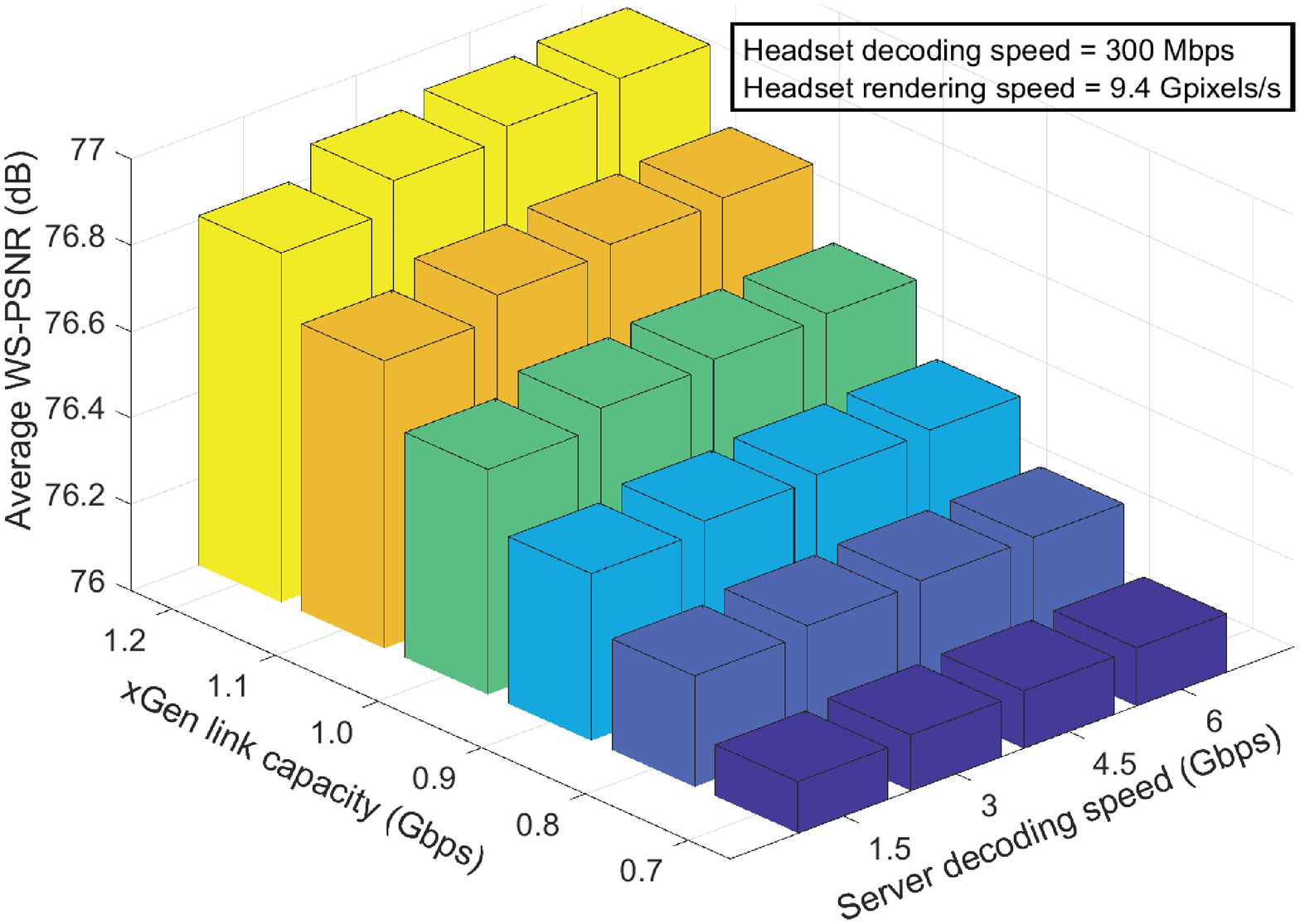}
\end{subfigure}%
\begin{subfigure}{.5\columnwidth}
  \centering
  \includegraphics[width=1\linewidth]{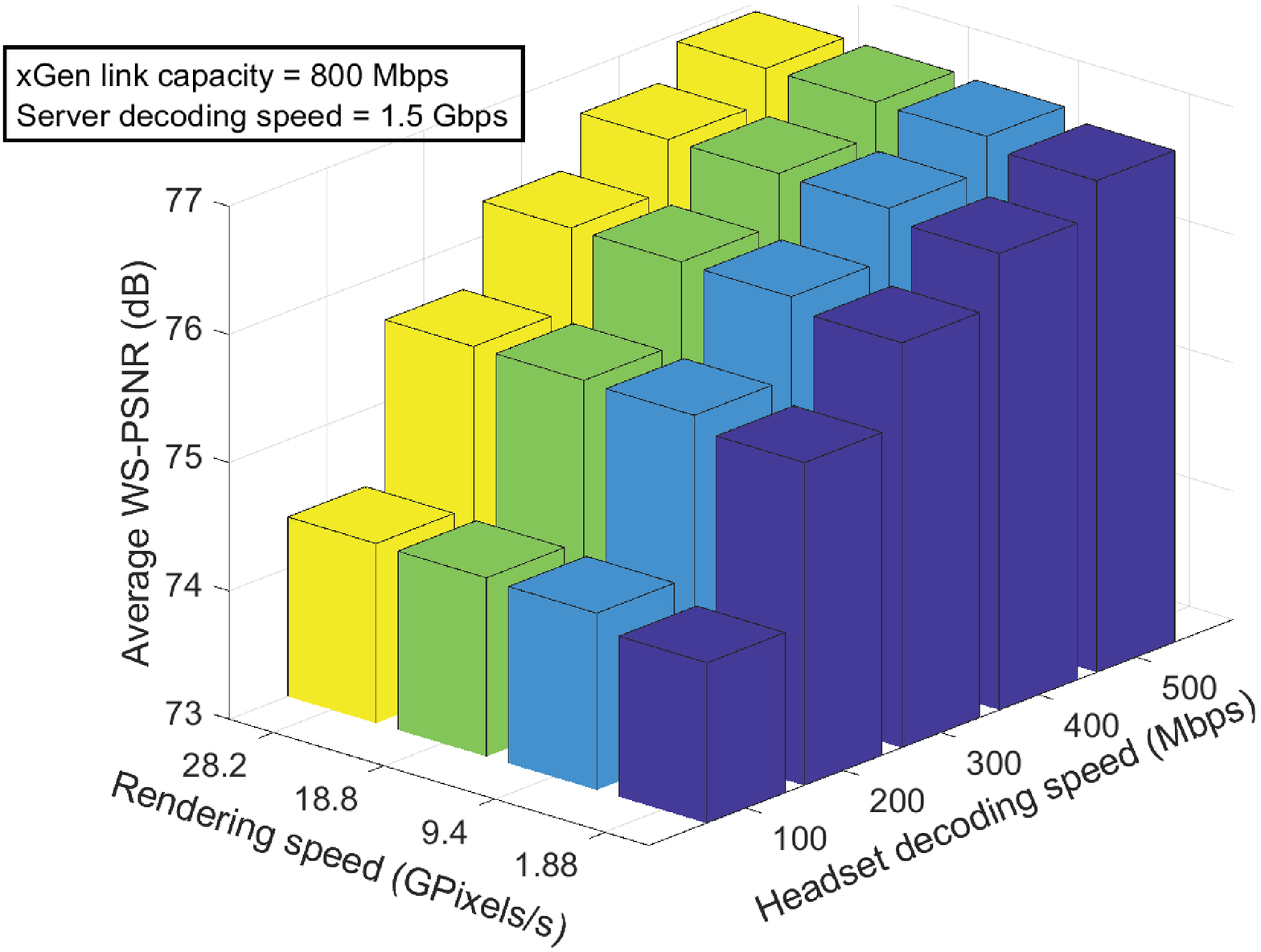}
\end{subfigure}
\vspace{-0.3cm}
\caption{Performance trade-offs: xGen transmission capability vs. edge server computing capability (left) and headset rendering vs. computing capabilities (right).}
\label{fig:PerformanceAnalysisTradeOffsCase1Case2}
\vspace{-0.3cm}
\end{figure}



{\em (i) 3-DOF \tsd video.} The performance results presented here are obtained by running the multi-user system optimization as described in the respective section, for many different placement configurations of the (static) users across the spatial area of the arena, and then computing the expected per-user performance. We recall that the users are not moving spatially across the arena during the session given the 3-DOF nature of the content. In Figure~\ref{fig:PerformanceAnalysisTradeOffsCase1Case2} (left), we examine the trade-offs between the transmission capacity of the xGen link to a user and the edge server's computing capability, and the resulting viewport quality experienced by the user. We can see that for lower xGen link capacities, increasing the server's decoding speed does not improve the delivered quality of immersion. That is because in such cases, almost all GOP tiles comprising the enhancement content layer are streamed as compressed data. Thus, the server's computing capability does not have much impact in such settings. On the other hand, for higher xGen link capacity values, the number of GOP tiles selected to be decoded at the server and streamed as raw data increases. Simultaneously, with the increasing link capacity, there is more room now for the remaining compressed tiles of the enhancement content layer to be transmitted at higher data rates. Both of these aspects contribute to higher quality of immersion for the VR user under such settings.

In Figure~\ref{fig:PerformanceAnalysisTradeOffsCase1Case2} (right), we examine the trade-offs between the rendering capability and decoding capability of the user's VR headset, and their interrelated impact on the delivered viewport quality. One can observe that the viewport quality moderately improves for around 0.2 dB as the rendering speed increases from 1.88 Gpixels/s to 9.4 Gpixels/s. Further increasing the rendering speed does not impact the delivered immersion fidelity. On the other hand, one can observe that increasing the headset decoding speed impacts the WS-PSNR more significantly. Concretely, for any given rendering speed examined in the figure, an improvement of around 2.5 dB is achieved as the device's decoding speed is increased from 100 Mbps to 500 Mbps. That is because, as the decoding speed increases, the device can decode GOP tiles compressed and transmitted at higher data rates, without violating the end-to-end latency constraints. Thereby, an improvement in the delivered viewport quality is achieved.

\begin{figure}[htb]
\centering
\begin{subfigure}{.5\columnwidth}
  \centering
  \includegraphics[width=1\linewidth]{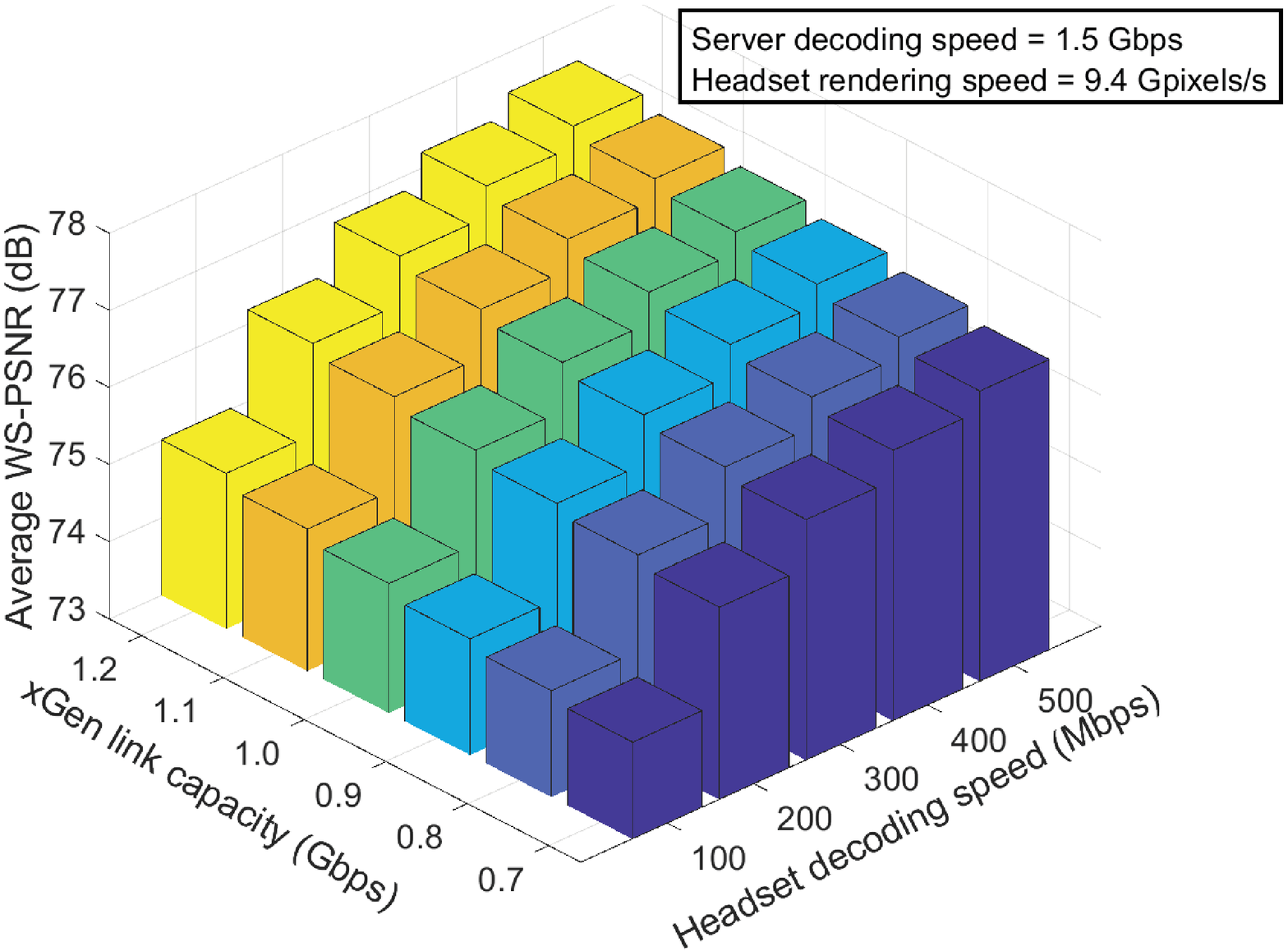}
\end{subfigure}%
\begin{subfigure}{.5\columnwidth}
  \centering
  \includegraphics[width=1\linewidth,trim={3cm 0 3cm 1cm},clip=true]{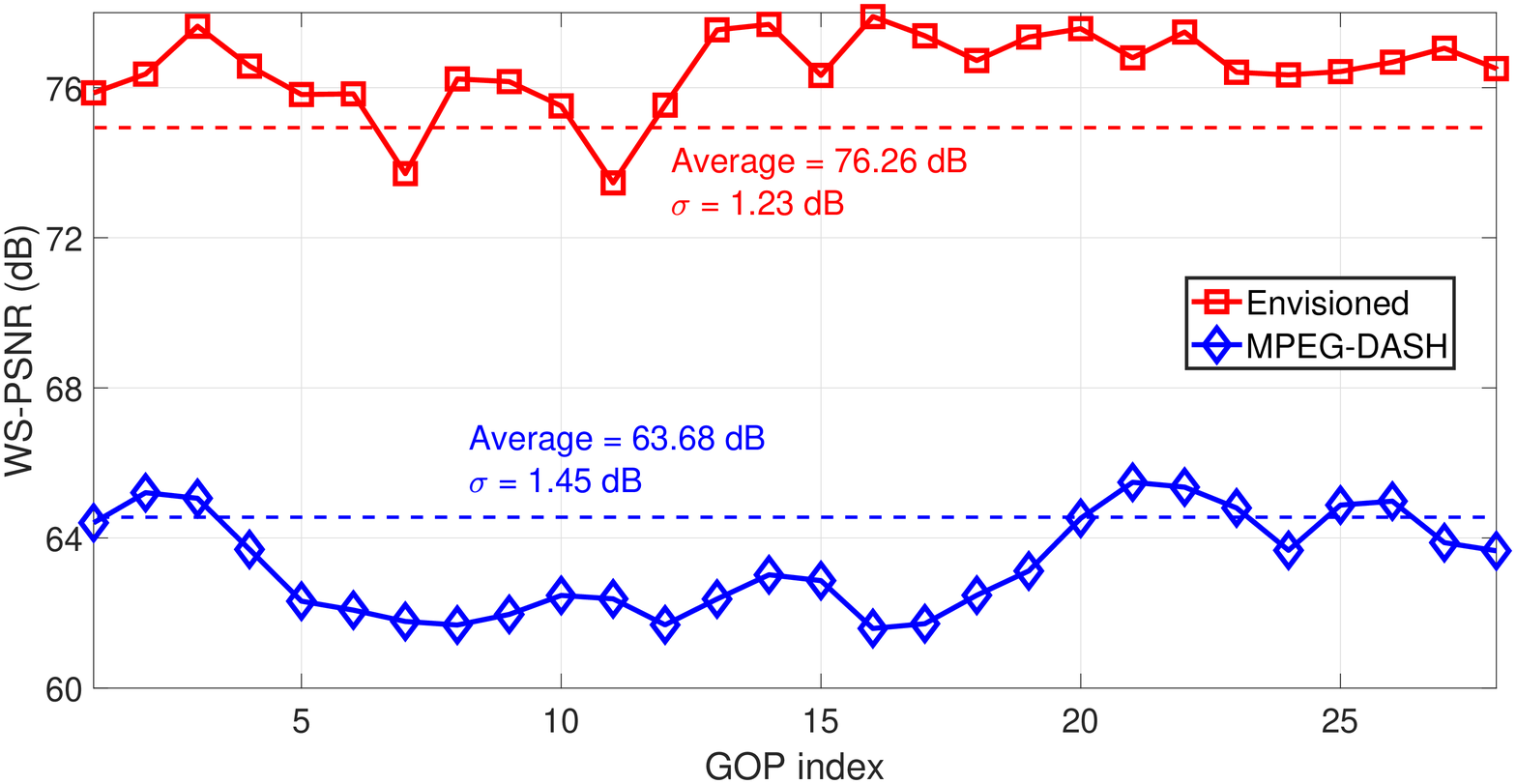}
\end{subfigure}
\vspace{-0.3cm}
\caption{Performance trade-offs: xGen transmission capability vs. client computing capability (left), and performance advances over a traditional state-of-the-art method (right).}
\label{fig:PerformanceAnalysisTradeOffsCase3Case4}
\vspace{-0.3cm}
\end{figure}

Next, in Figure~\ref{fig:PerformanceAnalysisTradeOffsCase3Case4} (left), we examine the trade-offs between the transmission capacity of the xGen link and the decoding speed of the headset, and their interrelated impact on the delivered quality of immersion. One can observe that the experienced WS-PSNR improves significantly as either the link capacity or the decoding speed increases. Concretely, the WS-PSNR increases for more than 1.5 dB when the link capacity increases from 700 Mbps to 1.2 Gbps. Similarly, when the link capacity is 700 Mbps, the WS-PSNR increases from 74.8 dB to 77.2 dB as the headset's decoding speed increases from 100 Mbps to 500 Mbps. Moreover, a WS-PSNR gain of around 4 dB is achieved when the link capacity is increased from 700 Mbps to 1.2 Gbps and the headset's decoding speed is increased from 100 Mbps to 500 Mbps. Increasing the value of either of these key capabilities enables GOP tiles compressed at higher data rates to be streamed from the server over the xGen link and decoded in time on the client device, thus improving the WS-PSNR significantly.

In Figure~\ref{fig:PerformanceAnalysisTradeOffsCase3Case4} (right), we examine the performance benefits enabled by the envisioned system over the traditional state-of-the-art. We implemented the current streaming standard MPEG-DASH following \cite{Petrangeli2017}, to stream the \tsd content, compressed using HEVC, over the traditional wireless connectivity link of a user. One can observe that significant performance gains in expected viewport quality and its variation are enabled. Concretely, MPEG-DASH can only achieve low average viewport quality of 63.68 dB, with a standard deviation of around 1.5 dB. These outcomes are in line with performance capabilities of emerging \tsd practices that can only stream lower resolution and lower frame rate \tsd videos at low to moderate quality at best, as outlined in the introduction of this article. On the other hand, the envisioned system enables high-quality viewport with expected WS-PSNR of 74 - 78 dB and standard deviation of 1.23 dB, for high frame rate and high resolution \tsd content, thus providing gains of 10-14 dB in immersion fidelity. These advances are motivating and are enabled by the dual connectivity streaming, scalable \tsd tiling, and edge computing that are synergistically integrated by the envisioned system together with rigorous end-to-end analysis, for maximum system performance efficiency.

%

A note on interpreting WS-PSNR results and their parallel with the more familiar PSNR metric may be appropriate here. The tile-level spherical distortion coefficients $w_{nm}$ are less than one and render the computed expected viewport distortion to be notably smaller. In our empirical evaluation, this has led to a consistent difference of around 11dB between the corresponding average PSNR and average WS-PSNR values. Moreover, a relative difference of a few dB in WS-PSNR may correspond to a difference of several dB in PSNR. Thus, the performance trade-offs of one method and its performance advances over another are highlighted more when interpreted through PSNR. Given the limited number of figures one can have, we opted to include the present collection of results under the WS-PNSR metric as most representative of the key performance aspects and advances introduced by the envisioned VR system concept. 
%
%
%
%
%
%
%

\begin{figure}[htb]
		\centering
		\includegraphics[width=0.9\linewidth]{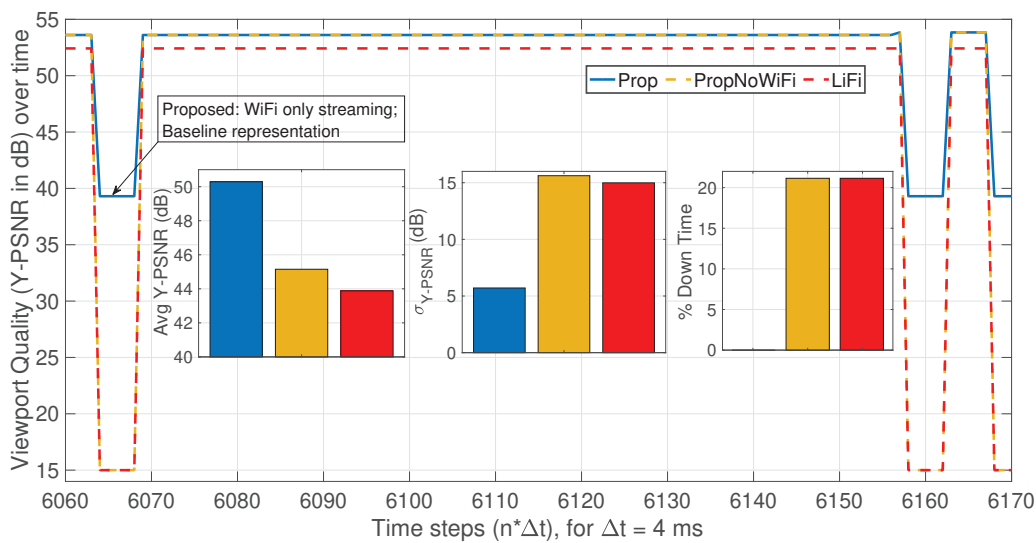}
		\caption{Sample temporal viewport quality in our system. (The user load is 6.) Y-PSNR: User viewport luminance (Y) video signal PSNR.}
		\label{fig:psnrtemporal}
		\vspace{-15 pt}	
	\end{figure}
	
\underline{\em (ii) 6DOF VR:} Finally, we highlight the simultaneous immersion quality and application reliability benefits of the envisioned VR system, when streaming challenging 6DOF VR content to mobile users in the arena. The performance results presented here are obtained by running the multi-user system optimization as described in the respective section. We focus on the use of VLC as a representative xGen technology in this performance analysis example. Concretely, we examine the delivered viewport quality over time for a user, in comparison to that for state-of-the-art reference methods. {\em LiFi} is a cellular VLC system that dynamically assigns a moving user to the cell and its stationary transmitter that maximize her SNR \cite{zeng2019angle}. We also examined a variant of our system that uses only (single-link) high-frequency wireless transmission (VLC) to stream the content. We refer to this variant later as PropNoWiFi, to indicate that no traditional (sub-6 GHz) wireless connectivity link is used in parallel in that case, in the proposed system. There are six simultaneous users in the arena. To contrast the earlier performance analysis examples, here we measured the traditional PSNR of the delivered viewport for a user, to asses its quality.

We can observe from Figure~\ref{fig:psnrtemporal} that the viewport quality varies over time for all three compared methods, due to one of the following two events, both induced from time to time by rapid head and body navigation movements. Either the VLC link is transiently dropped or a brief mismatch takes place between the viewport knowledge used to construct the enhancement content layer at the server and the actual user viewport at the receiving client (see Section~\ref{sec:Scal360TilingViewportAdaptation}: “Scalable \tsd Video Tiling and Viewport-Driven Adaptation”). Still, the observed viewport quality variation is much lower for our system, which considerably increases its quality of experience and reliability, relative to {\em PropNoWiFi} and {\em LiFi} that experience an application downtime during such instances. Concretely, viewport quality gains of 5dB and 7dB and three times smaller standard deviation of viewport quality are enabled over the latter two methods, as seen from the two smaller inset graphs to the left. We can also observe that the user experienced 22$\%$ application downtime for {\em PropNoWiFi} and {\em LiFi}, compared to 0$\%$ downtime for our method, as shown in the smaller inset graph to the right. These benefits merit the technical advances of the envisioned VR system.

	\begin{table}[htb]
		\caption{Average user performance (six users in the system).}
        \vspace{-15 pt}	
		\begin{center}
			\begin{tabular}{|c|c|c|c|c|}
				\hline
				\textbf{Method} & \textbf{Data rate (T)} & $\boldsymbol{\sigma}_T$ & \textbf{Y-PSNR} &  $\boldsymbol{\sigma}_{\textbf{PSNR}}$ \\
				\hline
				Prop. & 1020 Mbps & 164.83 Mbps & 50.37 dB & 5.79 dB \\
				\hline			
				{\em LiFi} & 550 Mbps & 210.89 Mbps & 43.83 dB & 14.96 dB \\
				\hline								
			\end{tabular}
			\label{tab:tab3}
		\end{center}
		\vspace{-15 pt}	
	\end{table}

In Table~\ref{tab:tab3}, we summarize the average user performance in regard to viewport quality and delivered data rate. We can see that our system consistently outperforms {\em LiFi} across all performance metrics considered, enabled by its synergistic technical advances. We highlight that MPEG-DASH could only deliver inadequate viewport quality of 39 dB in this setting, as expected. Finally, we note that the proposed system enables another performance benefit over LiFi, which is higher robustness to the user load, where a more graceful degradation in performance is provided, as the number of users in the arena is increased \cite{ChakareskiK:21}. These results cannot be included here due to the limited space.

	
\section{Conclusions and the road ahead}
The two pedagogical aspects of the article, to educate and inspire the reader, have been necessarily intertwined in its presentation, given the nature of its topic and its objectives. We started by providing an overview of virtual reality and its key present challenges towards further advancement, highlighting its prospectively most exciting use cases for our society in the future, comprising high-fidelity remote scene immersion and untethered lifelike navigation. Then, a broad survey of related studies using traditional approaches and synergistic advances in other technologies, and a system-level primer on virtual reality and \tsd video technologies were integrated, to put these challenges in perspective, highlight the shortcomings of present implementations, and inspire new approaches.
Next, the article provided a contextual high-level review of several emerging technologies and unconventional techniques, highlighting that only by their synergistic integration we can aim to overcome the bottlenecks of hyper-intensive computation, ultra-high data rate, and ultra-low latency toward the envisioned future societal applications.

A novel 6DOF VR system concept that embodies this integration in an indoor setting and a rigorous analysis that captures the fundamental interplay among communication, computation, and scalable signal representation that arises herein, and that optimizes the system's end-to-end performance were presented subsequently. Finally, several representative performance analysis examples were provided to highlight these trade-offs and the benefits of the envisioned system. These outcomes motivate the system as a broad research platform for further investigations spanning actual implementation and deployment, new analyses and optimizations, and integration of further technologies and techniques. Moreover, many of the advances introduced by this feature article and such follow-up work could benefit related technologies, such as augmented reality and holograms.

\section{Authors}

{\bf Jacob Chakareski} (\href{mailto:jacobcha@njit.edu}{jacobcha@njit.edu}) completed his Ph.D. degree in electrical and computer engineering at Rice University and Stanford University. He is an Associate Professor in the College of Computing at the New Jersey Institute for Technology (NJIT), Newark, New Jersey, 07103, USA, where he holds the Panasonic Chair of Sustainability and directs the AI-Enabled Laboratory for Virtual and Augmented Reality Immersive Communications and Network Systems. Dr. Chakareski organized the first National Science Foundation (NSF) visioning workshop on future virtual and augmented reality communications and network systems in 2018 \cite{Chakareski:19b}. He has held research appointments with Microsoft, HP Labs, and Ecole Polytechnique F\'{e}d\'{e}rale de Lausanne (EPFL), and served on the advisory board of Frame, Inc (acquired in 2019 by Nutanix, Inc.). His research interests span next generation virtual and augmented reality systems, UAV-IoT sensing and networking, fast reinforcement learning, 5G wireless edge computing and caching, ubiquitous immersive communication, and societal applications. He received the Adobe Data Science Faculty Research Award in 2017 and 2018, the Swiss NSF Career Award Ambizione (2009), the AFOSR Faculty Fellowship in 2016 and 2017, and Best Paper Awards at ICC 2017 and MMSys 2021. His research has been supported by the NSF, NIH, AFOSR, Adobe, Tencent Research, NVIDIA, and Microsoft. For further information, please visit \url{www.jakov.org}.

{\bf Mahmudur Khan} received the B.Sc. degree in Electrical and Electronic Engineering from Bangladesh University of Engineering and Technology, in 2011, the M.S. degree in CSE from the University of Nevada Reno, in 2015, and the Ph.D. degree in CE from the University of Central Florida, in 2018, respectively. He was a postdoctoral fellow at NJIT and is presently an Assistant Professor at York College. His research interests include free-space-optical communications, wireless ad hoc networks, and UAV communications.

{\bf Murat Yuksel} received the BS degree in CE from Ege University, Izmir, Turkey, in 1996, and the MS and PhD degrees in CS from RPI, in 1999 and 2002, respectively. He is an Associate Professor with the ECE Department, University of Central Florida (UCF), Orlando. Prior to UCF, he was with the CSE Department, University of Nevada, Reno as a faculty member until 2016. He worked as a software engineer with Pepperdata, Sunnyvale, California, and a visiting researcher with AT\&T Labs and the Los Alamos National Lab. His research interests include networked, wireless, and computer systems with a recent focus on big-data networking, UAV networks, optical wireless, and network management.

\section*{Acknowledgements}
To Ilija Chakareski and Stojan Angjelov for their love. The work of Jacob Chakareski and Mahmudur Khan has been supported in part by the National Science Foundation (NSF) under awards CCF-2031881, ECCS-2032387, CNS-2040088, CNS-2032033, and CNS-2106150; by the NIH under award R01EY030470; and by the Panasonic Chair of Sustainability at NJIT. The work of Murat Yuksel has been supported in part by the NSF under awards CNS-2115215, CNS-2120421, and CNS-1836741. The authors are grateful to the editors Dr. Matthew McKay, Dr. Robert Heath, and Dr. Laure Blanc-Féraud, as well as to the numerous anonymous reviewers for their constructive guidance and comments that have considerably helped improve the quality of this feature article.

\bibliographystyle{F://ThinkPad/Jakov/Publications/_StyleFiles/IEEEtran}
\def\baselinestretch{0.8}
\bibliography{F://ThinkPad/Jakov/Publications/_StyleFiles/myrefs}

\end{document}